\documentclass[pre,twocolumn,aps,showpacs,superscriptaddress]{revtex4}
\usepackage{graphics}
\usepackage{epsfig}
\usepackage{amsmath}
\usepackage{amssymb}
\usepackage{bbm}
\usepackage{bm}
\usepackage{color}
\usepackage{mathrsfs}

\newcommand{\pbar}{\langle p \rangle}

\newcommand{\sbar}{ \hat{\bar{\sigma}}}
\newcommand{\Sbar}{\hat{\cal S} }

\newcommand{\Abarp}{\langle a(p) \rangle}

\newcommand{\z }{\bar{z}}

\begin{document}
\title{Force rearrangements and statistics of hyperstatic granular force networks}

\author{Brian P.~Tighe}
\affiliation{Instituut-Lorentz, Universiteit Leiden, Postbus 9506, 2300 RA Leiden, The Netherlands}
\author{Thijs J.~H.~Vlugt}
\affiliation{Delft University of Technology, Process \& Energy Laboratory, Leeghwaterstraat 44, 2628 CA Delft, The Netherlands}

\date{\today}

\begin{abstract}

The heterogeneous force networks in static granular media --- formed from contact forces between grains and spanning from boundary to boundary in the packing --- are distinguished from other network structures in that they must satisfy constraints of mechanical equilibrium on every vertex/grain. Here we study the statistics of ensembles of hyperstatic frictionless force networks, which are composed of more forces than can be determined uniquely from force balance. Hyperstatic force networks possess degrees of freedom that rearrange one balanced network into another. We construct these rearrangements, count them, identify their elementary building blocks, and show that in two dimensions they are related via duality to so-called floppy modes, which play an important role in many other aspects of granular physics. We demonstrate that the number of rearrangements governs the macroscopic statistical properties of the ensemble, in particular the macroscopic flucutations of stress, which scale with distance to the isostatic point. We then show that a maximimum entropy postulate allows one to quantitatively capture many features of the microscopic statistics. Boundaries are shown to influence the statistics strongly: the probability distribution of large forces can have a qualitatively different form on the boundary and in the bulk. Finally, we consider the role of spatial correlations and dimension. All predictions are tested against highly accurate numerical simulations of the ensemble, performed using umbrella sampling.

\end{abstract}

\maketitle

Granular systems are athermal and dissipative: an undriven system will eventually reach a static, mechanically stable state and remain there. By repeatedly applying the same preparation protocol, whether numerically or experimentally, it is possible to build up an ensemble of static granular packings in which each element of the ensemble is a final state of the preparation protocol. Such an ensemble is very different from an equilibrium ensemble: not only is there no thermal equilibrium, there are no dynamics at all. 

Micromechanical models like soft spheres interacting via repulsive contact forces are an attractive way to study granular materials numerically; see, e.g., Ref.~\cite{vanhecke10} and references therein. Even at this level of abstraction, however, a theoretical description of the statistical or mechanical properties of disordered packings is daunting. In recent years, a model system called the force network ensemble (FNE) \cite{snoeijer04a} has proven to be an extremely useful tool for studying the disordered stress states of static granular packings. In its simplicity the FNE affords theoretical traction and permits highly accurate simulation, and yet it is detailed enough to reproduce many of the statistical and mechanical properties of numerical and experimental packings \cite{snoeijer04b,tighe05,ostojic06,ostojic06b,snoeijer06,vaneerd07,tighe08b}; Ref.~\cite{tighe10} provides a review. 

The force network ensemble is built on the observation that packings of noncohesive frictionless disks or spheres at finite pressure are {\em hyperstatic}: they possess a number of contacts in excess of that which would uniquely determine the contact forces from mechanical equilibrium. This means that for a single packing geometry there exist many different configurations of forces that satisfy force balance on each grain. This indeterminacy can, in principle, be lifted by specifying a contact force law, from which the forces can be determined given the grain positions. The conceit of the FNE, however, is {\em not} to specify this information, and instead to exploit the force indeterminacy to practical advantage. Because deformations are so small in packings of hard but not perfectly rigid grains, there is a strong separation between the grain scale and the contact scale. The idea is that, by averaging over all balanced force networks on a single frozen contact network, one captures fluctuations in the stresses that would also result from rearrangements of the grain positions. In the FNE, then, {\em grains} do not rearrange but {\em forces} do (Fig.~\ref{fig:illustration}). 

Here we describe the stress statistics of the force network ensemble within a statistical mechanics framework. This work represents an elaboration and significant expansion on recent results, which demonstrated that the statistics of local stresses within the FNE can be described using a maximum entropy principle \cite{tighe08b}. The notion of an ensemble description of static granular matter dates to Edwards, who proposed an ensemble of packings characterized by like boundary conditions \cite{edwards89}. Though conceptually appealing, the Edwards ensemble is difficult to probe theoretically. In this spirit the FNE can be seen as a restricted but more accessible version of Edwards' ensemble. In ensuing years, a number of authors have proposed granular ensembles in which stress plays a role similar to that of energy in an equilibrium ensemble \cite{evesque99,kruyt02,bagi03,ngan03,goddard04,henkes07,edwards08,metzger08}, and we shall see that the FNE naturally lends itself to such an approach.

This paper is divided in two main sections. Section \ref{sec:rearrangements} develops the FNE in greater detail, with particular emphasis on the {\em force rearrangements} that transform one force network into another. Force rearrangements can be constructed ``by hand'' by considering the local force balance constraints on grains in a static packing. Their onstruction allows us to motivate the ensemble differently than previously \cite{snoeijer04a,tighe10}, in a manner that anticipates the theoretical framework to follow. Whereas in prior work a constraint on the global stress was explicitly imposed as a separate constraint, we demonstrate that the ensemble, defined in terms of force rearrangements, can be built up in a manner that obviates this step. It turns out that the force rearrangements act in such a way as to leave the global stress unchanged, {\em without} imposing an external constraint. This leads naturally to the idea of the FNE as an analog of the microcanonical ensemble: it is a system with ``dynamics'', here realized by the force rearrangements, that leave a quantity invariant. We will demonstrate that in the FNE that quantity is stress or a related invariant, rather than energy. 

Given its similarity to the equilibrium microcanonical ensemble, it is natural to anticipate that the ensemble dynamics maximize an entropy. In Section \ref{sec:statmech} we write down this entropy and develop a framework that allows for description of the statistics of macroscopic stresses in the ensemble: We derive the FNE equation of state and an expression for macroscopic pressure fluctuations in the canonical FNE. We then turn to the statistics of local measures of the stress, particularly pressure $p$ at the grain scale; our focus is on the form of the pressure probability distribution function of $p$ for asymptotically small and large $p$. We also treat spatial correlations and dimensions $d \ge 3$, respectively. Finally, Section \ref{sec:outlook} gives a discussion and future outlook.

\section{Force rearrangements and the FNE}
\label{sec:rearrangements}

Energy ceases to be an invariant of the dynamics in a dissipative system. In ensembles of static granular packings, moreover, we explicitly decline to consider the dynamical states that the system traverses when it passes from one static state to another. In this context the force network ensemble can be understood as an intermediate case. As described above, each force network in the ensemble is meant to represent the stress state of a static packing. At the same time, we shall see that the ensemble {\em does} have a discrete dynamics, at least in the sense of Monte Carlo rearrangements of the forces. In this section we will construct these force rearrangements for arbitrary disk packings and then use them to give a more precise definition of the force network ensemble. We then demonstrate that the Monte Carlo dynamics of the FNE leave two related quantities invariant. These quantities are then natural candidates on which to base a statistical description; this is the topic of Section \ref{sec:statmech}.

\begin{figure}[tbp] 
\centering
\includegraphics[clip,width=0.9\linewidth]{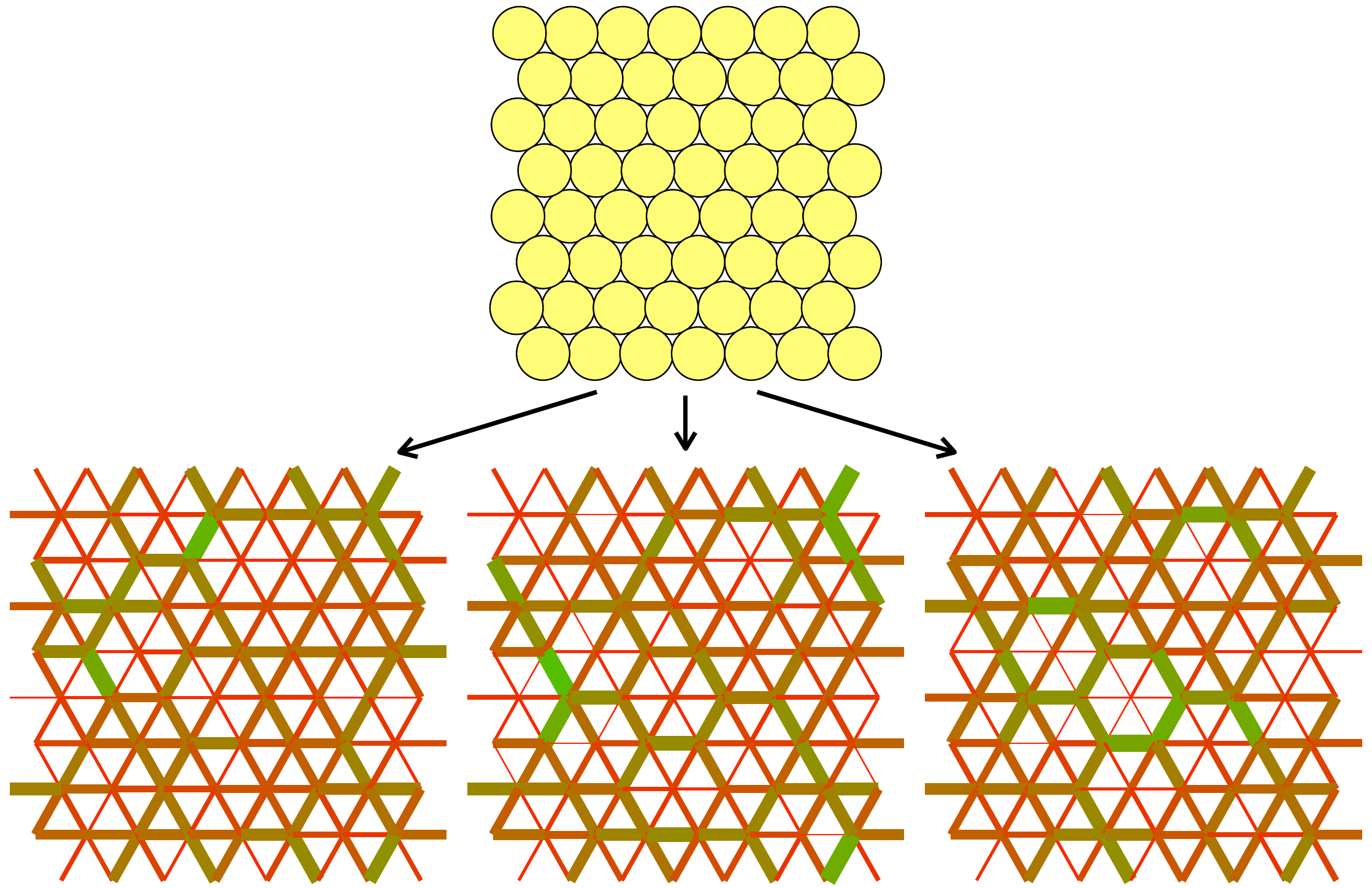}\\
\caption{A single hyperstatic contact network, such as the periodic frictionless triangular lattice, admits many force networks in which forces balance on every grain; here three are shown. Lines represent contact forces, with their thickness proportional to the force magnitude. Though the contact network is ordered, typical force networks are disordered. }
\label{fig:illustration}
\end{figure}

To begin, we illustrate the concept of force rearrangements in a particularly simple contact network, the triangular lattice (Fig.~\ref{fig:illustration}). Although the contact network is ordered, it admits many different force balanced configurations of the contact forces between the grains, the overwhelming majority of which are disordered. This is because the triangular lattice is strongly hyperstatic. 

Hyperstaticity can be quantified in terms of the mean coordination number of a packing $\z$; in the triangular lattice every grain has six contacts, and $\z = 6$. A generic disk or sphere packing must have a minimum coordination $z_{\rm iso}$ to satisfy force balance on each grain. To see this, one can think of the individual contact forces as degrees of freedom. There must be enough degrees of freedom to satisfy all the constraints of mechanical equilibrium. Let us consider frictionless spheres, for which all contact forces have just one component, a normal force, and hence torque balance is satisfied automatically. In a frictionless packing of $N$ grains in $d$ dimensions, these constraints are the equations of force balance, of which there are $N_{\rm fb} = dN$. There is one contact force for each of the $N_{\rm c} = \frac{1}{2}\z N$ contacts. A system is called {\em isostatic} when there are just enough forces to satisfy the constraints. Requiring $N_{\rm fb} = N_{\rm c}$ yields $z_{\rm iso} = 2d$. The calculation can be repeated for frictional spheres, in which case there are both additional constraints due to torque balance and additional degrees of freedom due to the tangential force components at each contact. The bulk of this work is dedicated to frictionless packings, although many results have straightforward generalizations to the frictional case.

\subsection{The triangular lattice}

\begin{figure}[tbp] 
\centering
\includegraphics[clip,width=0.35\linewidth]{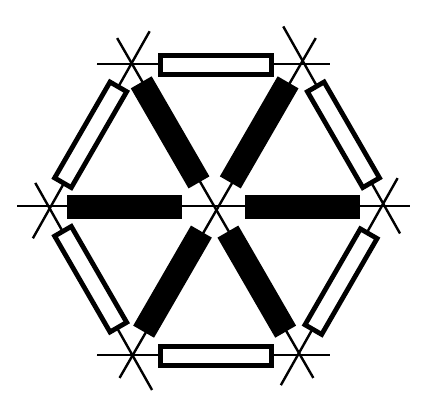}\\
\caption{Monte Carlo or ``wheel'' move in the frictionless triangular lattice \cite{tighe05}. Force is added to spoke contacts (solid bars) and subtracted from six rim contacts (open bars), or vice versa. The net vector force on 
each of the seven participating grains is unchanged.}
\label{fig:wheel}
\end{figure}

The frictionless triangular lattice has an excess coordination $\Delta z := \z - z_{\rm iso} = 2$. Because each contact is shared by two grains, this means that there is $\frac{1}{2} \Delta z = 1$ degree of freedom per grain. Because the triangular lattice is ordered, it is possible to construct this degree of freedom by inspection. It is shown in Fig.~\ref{fig:wheel}, and known as a ``wheel move'' due to its appearance \cite{tighe05}. A wheel move is a rearrangement of 12 forces: six ``spoke forces'', which are changes to the contact forces on a central grain, and six ``rim forces'', which change the contact forces between the nearest neighbor grains. The spoke and rim forces have equal changes in magnitude but opposite sign. By design, all seven grains are in force balance. Because the equations of force balance are linear, a wheel move can be added to any balanced force network on the triangular lattice, such as the ones in Fig.~\ref{fig:illustration}, without violating force balance. 

By labeling each contact force ${\vec f}_{ij} = -{\vec f}_{ji}$ by the grains it acts on, we can construct a vector ${\bf f} = \lbrace {\vec f}_{ij} \rbrace$ containing all the unique forces in the force network. One possible force balanced configuration in the frictionless triangular lattice is the force network ${\bf f}_0$ in which every force has the same magnitude $\bar{f}$. In the same notation, we label the $N$ wheel moves, one for each grain, as $\lbrace \delta {\bf f}_k \rbrace$, where the index indicates the grain on which they are centered. 

We now define the the (isotropic) force network ensemble on the frictionless triangular lattice to be all noncohesive force networks $\lbrace {\bf f}_k \rbrace$ that can be reached by applying any combination of wheel moves to the force network ${\bf f}_0$. That is, a force network in the FNE can be expressed
\begin{equation}
{\bf f} = {\bf f}_0 + \sum_{k=1}^{N-1} w_k \, \delta {\bf f}_k \,.
\label{eqn:fnetri}
\end{equation}
The weights $\lbrace w_k \rbrace$ then serve as coordinates of the force network ${\bf f}$ in a high-dimensional space.
Only $N-1$ wheel moves are required for a linearly independent set \cite{tighe05}, hence the upper bound in the sum. To see this, consider the application of every wheel move with equal weight -- the changes to each force sum to zero. Therefore the $\lbrace w_k \rbrace$ are only defined modulo their sum or, equivalently, one of the rearrangements is dictated by the sum of the others.  

The restriction to {\em noncohesive} forces means that the weights $\lbrace w_k \rbrace$ must be chosen so that the normal force component $(f_{\rm n})_{ij} \ge 0$ for all contacts; recall that in frictionless sphere packings all forces are normal forces. From the linearity of the constraints it follows that the  space of force networks is a convex polytope \cite{mcnamara04,snoeijer04b,unger05}. Therefore the choice of ${\bf f}_0$ is not essential: replacing ${\bf f}_0$ by any force network ${\bf f}_0^\prime$ that can be reached by applying wheel moves to ${\bf f}_0$ does not change the ensemble defined according to Eq.~(\ref{eqn:fnetri}) \footnote{We exclude using networks {\em on the boundary} of the polytope for ${\bf f}_0$, which can cause difficulties with sampling. See Ref.~\cite{tighe05} for a discussion of this point.}. We show in Section \ref{sec:invariants} that the reference network ${\bf f}_0$ serves to select the stress tensor $\sbar$. 

Dynamics in the FNE can be visualized as a random walk in the space of force networks. As their name anticipates, the wheel moves serve as Monte Carlo moves. To sample the space of force networks, Monte Carlo moves are randomly selected and used to update the current force network .
The size of the move is uniformly selected from the interval of possible step sizes, determined by the positivity constraint on the affected contact forces \footnote{
Other sampling methods are possible. E.g.~when employing umbrella sampling (see Appendix C) moves are accepted/rejected according to Eq.~(\ref{eqn:A3}).}. 
In this way, for sufficiently long runtimes, the space of force networks is sampled with flat measure; see Refs.~\cite{vaneerd07} and \cite{tighe05} for details. 

We now turn to constructing force rearrangements in disordered contact networks. With the exception of Section \ref{sec:3d}, this work is concerned primarily with two-dimensional systems.  We therefore illustrate the construction of force rearrangements in two-dimensional contact networks, where we have access to a geometric construct called a Maxwell-Cremona diagram or reciprocal tiling \cite{maxwell1864}. The reciprocal tiling is a helpful tool and, furthermore, allows us to make connections to the concept of {\em floppy modes}, which play an important role in other aspects of granular physics \cite{tkachenko99,wyart05,vanhecke10}. We therefore describe the reciprocal tiling first before addressing disordered force rearrangements. We emphasize, however, that force rearrangements can also be constructed in higher dimensions \cite{vaneerd07,vaneerd09}.

\subsection{Maxwell-Cremona diagrams}
\begin{figure}[tbp] 
\centering
\includegraphics[clip,width=0.85\linewidth]{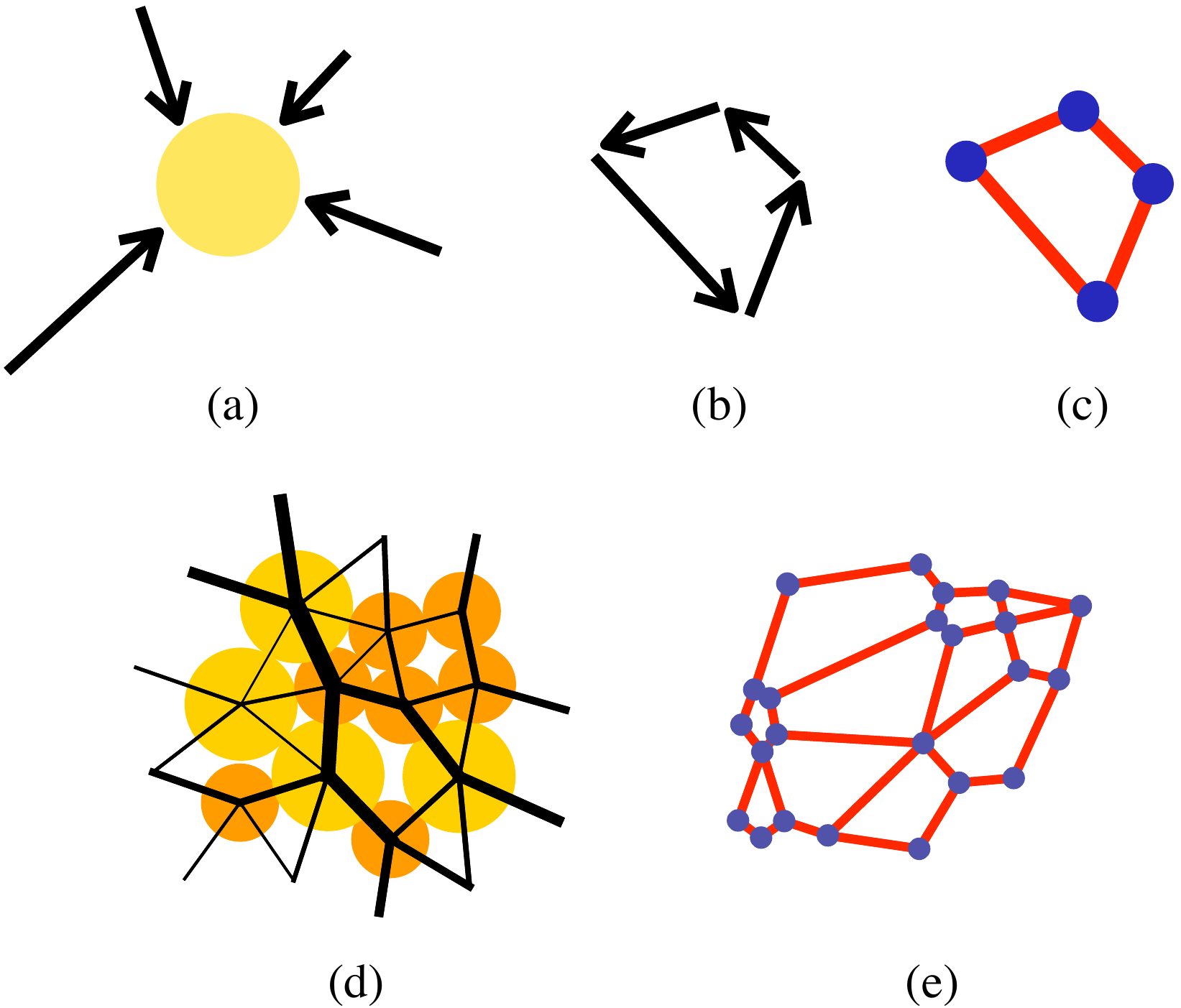}\\
\caption{Constructing a tile in the Maxwell-Cremona diagram, or reciprocal tiling. (a) A disk with four contacts. Vector contact forces imposed by the neighboring disks are indicated by arrows. Note that the forces need not be frictionless, i.e.~need not be parallel to the segment connecting the contacting disks' centers. (b) Each contact force is rotated by $\frac{\pi}{2}$ and drawn end-to-end, proceeding around the grain in a right-hand fashion. (c) The polygon enclosed by the vectors is the grain's corresponding tile in the tiling. The corners of the tile are vertices in the Maxwell-Cremona diagram. (d,e) Due to Newton's third law, tiles corresponding to contacting grains can be placed flush against one another. The result is a tiling or tesselation. To determine the area of the tiling, it suffices to know the boundary forces on the packing; alternatively, one can sum the areas of individual tiles.}
\label{fig:arbgrain}
\end{figure}

Contact forces in a packing may be used to construct a Maxwell-Cremona diagram, in which pairs of action-reaction forces between grains are mapped to a tiling of the plane. An edges in the tiling has a length proportional to the magnitude of the corresponding force, and its orientation is perpendicular to the vector force. This is most easily seen graphically (Fig.~\ref{fig:arbgrain}a-c): the boundary of a tile is constructed by rotating the vector forces acting on a grain by $\pi/2$ and placing them end to end in a right-hand fashion. Because the boundary is the vector sum of the contact forces acting on the grain, the boundary is closed (a polygon) whenever the grain is in force balance and not subject to body forces. Though the tiling can be generalized to incorporate body forces \cite{tighe09}, they will not be considered here.

By Newton's third law, tiles of contacting grains have faces of like length and orientation. Hence tiles may be placed next to each other seamlessly, and the Maxwell-Cremona diagram is built up tile-by-tile in this fashion. For a packing in static force balance subject to imposed forces at the boundary, the corresponding Maxwell-Cremona diagram has a closed boundary and no internal gaps, as illustrated in Fig.~\ref{fig:arbgrain}d and e. For a periodic packing, the tiling is also periodic. The reciprocal space coordinates of the diagram's vertices are, after rotation by $-\pi/2$, the void forces of Satake \cite{satake93} and equivalently the loop forces of Ball and Blumenfeld \cite{blumenfeld02}. 

The reciprocal tiling exists as a consequence of static force balance in the packing. The construction makes no assumptions regarding the presence or absence of tangential or tensile forces, and torque balance is not a necessary condition for its existence. Nevertheless, all the force networks we study here are noncohesive and (trivially) torque balanced.

For later convenience let us now calculate the mean coordination number in a periodic reciprocal tiling. For each grain, contact, and void in the packing there is a corresponding tile, edge, and vertex, respectively, in the tiling, and the topology of the tiling is the same as that of the dual graph of the contact network \cite{graphtheory}. Euler's relation for a periodic network relates the number of grains, contacts, and voids in a packing:
\begin{equation}
N - N_c + \widetilde{N} = 0 \,,
\label{eqn:euler}
\end{equation}
where $\widetilde{N}$ is the number of voids, or equivalently the number of vertices in the reciprocal tiling. The mean coordination number in the tiling is $\tilde{z} = 2N_c/\widetilde{N}$. Combined with Eq.~(\ref{eqn:euler}) this gives
\begin{equation}
 \frac{1}{2} = \frac{1}{\z} + \frac{1}{\tilde{z}} \,,
\label{eqn:dualz}
\end{equation}
where we have written the relation in a form that emphasizes the symmetry between $\z$ and $\tilde{z}$.
In particular, whenever $\z > z_{\rm iso}$, $ \tilde{z}  < z_{\rm iso}$.

\subsection{Floppy motions and force rearrangements}

\begin{figure}[tbp] 
\centering
\includegraphics[clip,width=0.55\linewidth]{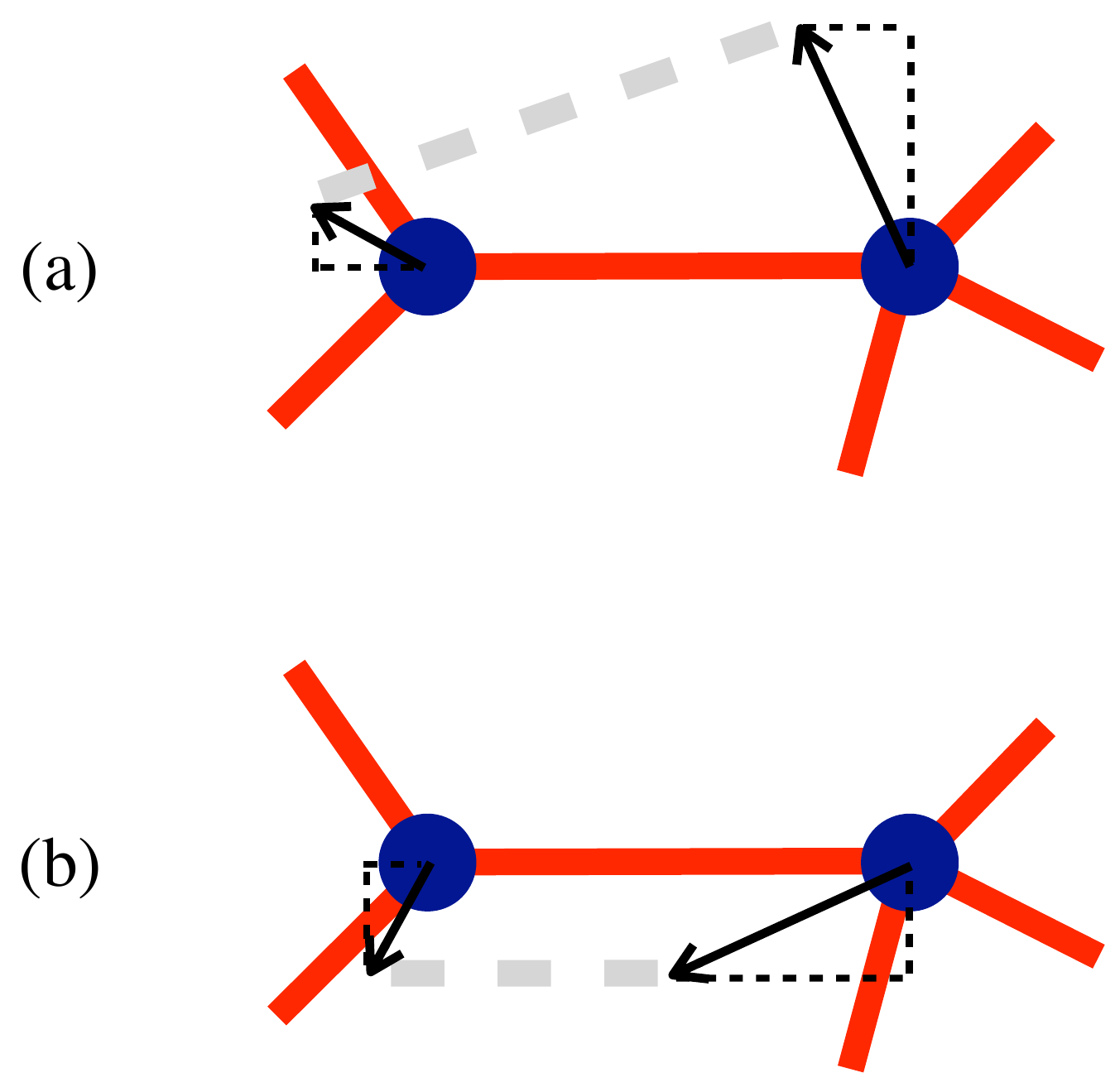}\\
\caption{Two vertices in a reciprocal tiling. (a) Floppy motion displacements (arrows) may rotate the edge joining the vertices (thick dashed line) but not change its length. (b) Rotating the displacements from (a) by $\pi/2$ produces a motion that changes the edge's length, but not its orientation. }
\label{fig:floppy2force}
\end{figure}

We now show that there is an intimate connection between the force rearrangements in frictionless packings and floppy modes --- non-rigid body motions of a collection of particles that do not change the inter-particle distances. The connection enters via the Maxwell-Cremona diagram. 
A motion of the tiling vertices that does not change the distance between vertices is a floppy mode.
Let us label the coordinates of vertices {\em in the reciprocal tiling} as $\lbrace {\vec h}_i \rbrace$, $i = 1 \ldots \widetilde{N}$.
If there is a floppy mode in the tiling, it must be that for each vertex with coordinate ${\vec h}_i$ connected by an edge to a vertex at ${\vec h}_j$, their motions $\delta{\vec h}_i$ and $\delta{\vec h}_j$ are such that
\begin{equation}
({\vec h}_i - {\vec h}_j) \cdot (\delta{\vec h}_i - \delta{\vec h}_j) =0\,.
\label{eqn:floppy}
\end{equation}
That is, to leading order in the motions, the relative motion of the two vertices can have no component along the original orientation of the edge between them; this is illustrated in Fig.~\ref{fig:floppy2force}a. The same consideration applies to every edge in the tiling.

The key observation is as follows, and is illustrated in  Fig.~\ref{fig:floppy2force}. A given floppy mode describes a set of vertex motions $\lbrace \delta {\vec h}_i \rbrace$. We take these vertex motions, rotate each of  them by $\pi/2$, and label the new motions $\lbrace \delta {\vec h}_i^\perp \rbrace$. By Eq.~(\ref{eqn:floppy}), the new relative motions are such that, to leading order, the edge between any pair of connected vertices does not rotate; instead it translates and changes its length. This is most easily seen in Fig.~\ref{fig:floppy2force}. Recall, however, that the edges in the reciprocal tiling are the forces in the original packing, and that a force rearrangement in a frictionless packing must change the magnitude but not the orientation of a contact force. Moreover, because the motions are performed in the reciprocal tiling, which only exists if the packing is in force balance, they must respect force balance. Therefore floppy modes in the reciprocal tiling can be mapped to force rearrangements in the packing simply by rotating their motions by $\pi/2$. We will use this fact to construct a set of ``disordered wheel moves''.

\subsection{Rearrangements in triangulations}
\label{sec:triangulations}

\begin{figure}[tbp] 
\centering
\includegraphics[clip,width=0.35\linewidth]{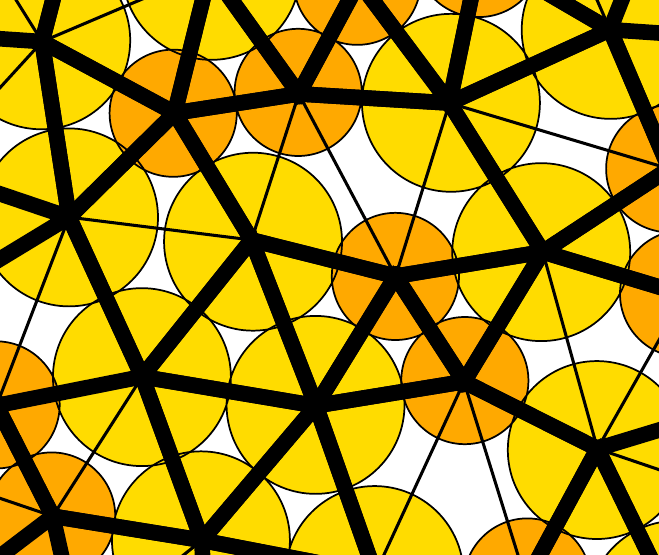}
\caption{A portion of a soft disk packing; lines indicate edges in the Delaunay triangulation of the disk centers. Edges with thick lines are also a part of the packing's contact network. }
\label{fig:dt}
\end{figure}

\begin{figure}[tbp] 
\centering
\includegraphics[clip,width=0.9\linewidth]{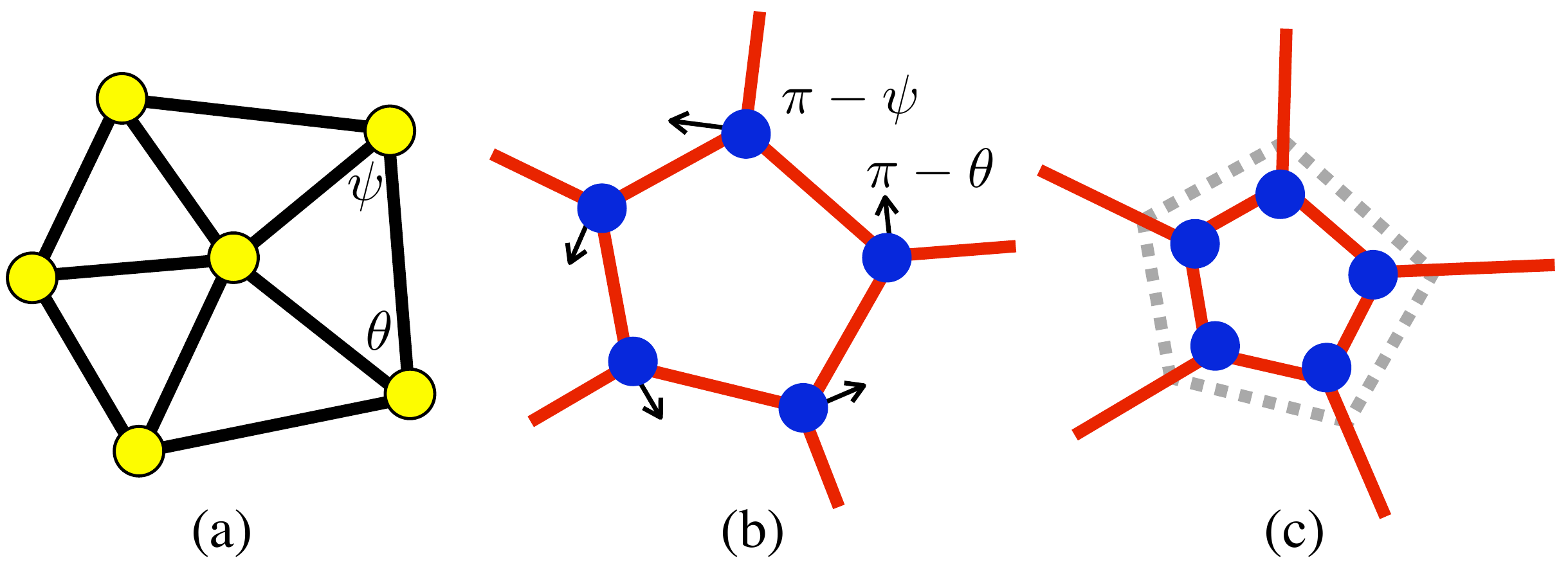}
\caption{Disordered wheel moves, the building blocks of frictionless force rearrangements. (a) In a triangulation, every grain/vertex is surrounded by a closed loop formed by its nearest neighbors. We define angles $\theta$ and $\psi$ for each triangle around the grain. (b) The grain's reciprocal tile has just one dangling edge from each vertex. A local floppy motion can be implemented by moving each vertex of the tile perpendicular to the dangling edge with a magnitude $\delta h_i$ determined in the text. (c) The disordered wheel move can be constructed by rotating the floppy motions $\lbrace \delta{\vec h}_i \rbrace$ by $\frac{\pi}{2}$. Moves in contact networks that are not triangulations can be constructed from linear combinations of these wheel moves. }
\label{fig:floppy}
\end{figure}

We will construct force rearrangements in disordered contact networks in two steps. In the first step we consider contact networks that are periodic triangulations of the plane. There it is possible to identify straightforward generalizations of the wheel moves in the triangular lattice. In the second step we construct linear combinations of these local rearrangements to form force rearrangements on contact networks that are not triangulations.

Given the grain positions in a disk packing, the packing's Delaunay triangulation can be constructed \cite{voronoi}; see Fig.~\ref{fig:dt} for an example. A Delaunay triangulation connects disks that are geometric neighbors, regardless of whether they are actually in contact. 
In systems with large polydispersity there may be physical contacts that are not Delaunay edges; it is possible to correct for this, but here we assume that all physical contacts are indeed edges in the triangulation.
An important feature of triangulations for present purposes is the following: each vertex is enclosed by a loop formed from edges connecting the neighbors of the vertex, i.e.~the vertices to which the central vertex is connected by an edge. This is depicted in Fig.~\ref{fig:floppy}a. 
The idea is to first treat all edges in the triangulation as if they can carry force. We then identify a set of floppy modes centered on each tile in the tiling; the floppy modes then give the disordered analog of a wheel move by the above prescription. In general, the wheel moves will make changes to the force on edges that are part of the Delaunay triangulation but not the contact network. To correct for this, we later construct linear combinations of the wheel moves that do not alter the force on these ``deleted edges''.

Fig.~\ref{fig:floppy}b depicts the reciprocal tile corresponding to the grain in Fig.~\ref{fig:floppy}a. Because the neighboring grains form a closed loop about the central grain, the central grain's corresponding tile has just one edge extending from each of its vertices. These ``dangling edges'' allow a floppy mode to be constructed by displacing the vertices of just a single tile. For a tile with $z_i$ vertices, each with coordinate ${\vec h}_j$ in the reciprocal space, we prescribe displacements $\lbrace \delta{\vec h}_j \rbrace$, $j=1\ldots z_i$. Each $\delta {\vec h}$ must be orthogonal to the dangling edge at its vertex, in accordance with Eq.~(\ref{eqn:floppy}). The magnitudes $\lbrace \delta{h}_j \rbrace$ of the vertex motions can be determined from the constraints that the lengths of the tile's edges remain unchanged to leading order in the motions. Although there are $z_i$ magnitudes and as many constraints, this is a one parameter family of motions: the constraint equations are homogeneous in the magnitudes $\lbrace \delta h_i \rbrace$, hence there is a degeneracy among them.

To see the degeneracy, we apply Eq.~(\ref{eqn:floppy}) to each edge around the boundary of the tile. This gives $\delta h_j \sin{\theta_j} = \delta h_{j+1} \sin{\psi_j}$, where the angles $\theta$ and $\psi$ are defined in Fig.~\ref{fig:floppy} and the index $j$ increments around the tile vertices in a righthand fashion, modulo $z_i$. Taking the product of this relation applied to each edge of the tile, one finds
\begin{equation} \label{eqn:degeneracy}
\prod_{i=1}^z \sin{\theta_i} = \prod_{i=1}^z \sin{\psi_i} \,.
\end{equation}
Note that Eq.~(\ref{eqn:degeneracy}) involves only angles; the magnitudes of the vertex displacements have dropped out. In fact, the same relation arises by repeatedly applying the law of sines to the triangles surrounding the central node in the original triangulation (Fig.~\ref{fig:floppy}a). This is a consequence of the fact that edges between nearest neighbors of the central grain form a closed loop. It means that Eq.~(\ref{eqn:degeneracy}) is satisfied automaticallly by geometry, and only $z_i-1$ of the $z_i$ apparent constraints on the magnitudes $\lbrace \delta{h}_i \rbrace$ are independent. The resulting one parameter family of motions is a floppy mode in the Maxwell-Cremona diagram; it is, by construction, localized to the vertices of a single tile. By following the prescription $\lbrace \delta{\vec h}_i \rbrace \rightarrow \lbrace \delta{\vec h}^\perp_i \rbrace$ we have thus constructed a localized force rearrangement (Fig.~\ref{fig:floppy}c). There is one per grain; we call them disordered wheel moves and label them $\lbrace \delta {\bf f}_k^\mathrm{local} \rbrace$, $k = 1 \ldots N$.

The force network ensemble on a triangulation can thus be defined to comprise all force networks of the form
\begin{equation}
{\bf f} = {\bf f}_0 + \sum_{k=1}^{N-1} w_k\,\delta{\bf f}^{\rm local}_k \,.
\end{equation}
The weights of the disordered wheel moves $\lbrace w_k \rbrace$ again serve as coordinates of the state in the space of force networks. The sum runs to $N-1$ because, as in the triangular lattice, there is a sum rule on the disordered wheel move weights (see Appendix A).
The particular solution ${\bf f}_0$ can be identified via simulated annealing \cite{snoeijer04b} or linear programming \cite{tighe05}.

The reciprocal tiling is similarly useful in constructing the counterpart of disordered wheel moves in frictional systems. This procedure is described in Appendix B.

\subsection{Rearrangements in general disordered contact networks}
\label{sec:extendedmoves}

Let us now consider rearrangements in periodic contact networks that are not triangulations. Euler's relation, Eq.~(\ref{eqn:euler}), guarantees that a packing's Delaunay triangulation has mean coordination number $z_{\rm tri} = 6$.  By comparison,
the physical contact network with mean coordination number $\z$ has only $N_{\rm c} = \frac{1}{2}\z N$ contacts, so that there are $N_{\rm d} = 3N - N_{\rm c}$ deleted edges in the triangulation, i.e.~edges in the triangulation that cannot transmit force because they do not correspond to contacts in the packing. Therefore we must impose $N_{\rm d}$ constraints on the disordered wheel moves from the triangulation: they must always act in combinations chosen so as to neither add nor subtract force on deleted edges. We now give a prescription for constructing these combinations.

Define $\bf F$ to be the $N-1 \times \frac{1}{2}\z N$ matrix with $\delta {\bf f}_k^\mathrm{local}$, $k = 1 \ldots N-1$, as its $k^\mathrm{th}$ column. Likewise define $\bf D$ to be a $\frac{1}{2}\z N \times N_\mathrm{d}$ matrix such that each column has a unit entry for a unique deleted edge, and all other elements zero. Then, for ${\bf w} = \lbrace w_k \rbrace$, the matrix operation $({\bf D}{\bf F}) \,{\bf w}$ returns the effect of a superposition of local moves with weights ${\bf w}$ on the deleted edges. As the deleted edges cannot carry force, this must be null:
\begin{equation}
({\bf D}{\bf F})\,{\bf w} = 0 \,.
\end{equation}
A set of vectors $\lbrace {\bf w}_n \rbrace$, $n = 1 \ldots N_\mathrm{w}$, spanning the null space of $\bf D  F$ gives a basis set of linear superpositions of the disordered wheel moves $\lbrace \delta {\bf f}_k^\mathrm{local} \rbrace$ having null effect on the deleted edges. These are our force rearrangements. Assuming none of the deleted edges are redundant, which is generally true for a disordered network, each deleted edge reduces the number of force rearrangements by one from the $N-1$ independent disordered wheel moves available in a triangulation: $N_{\rm w} = N-N_{\rm d} -1$, or
\begin{equation}
\label{eqn:Nw}
N_{\rm w} = \frac{1}{2} \Delta z \, N - 1 \,.
\end{equation}
Each such rearrangement can be written
\begin{equation}
\delta{\bf f}_n^{\rm ext} = \sum_{k=1}^{N-1} [{\bf w}_n]_{k} \,\delta{\bf f}_k^{\rm local} \,,
\end{equation}
where $[{\bf w}_n]_{k}$ is the $k^{\rm th}$ element of ${\bf w}_n$. The label emphasizes that these rearrangements are spatially extended, being composed of multiple disordered wheel moves.

In analogy to Eq.~(\ref{eqn:fnetri}), we can now define the FNE on a disordered contact network as the set of force networks expressible as
\begin{equation}
{\bf f} = {\bf f}_0 + \sum_{n=1}^{N_{\rm w}} c_n \,\delta{\bf f}_n^{\rm ext}\,,
\label{eqn:genfne}
\end{equation}
where the $N_{\rm w}$ coefficients $\lbrace c_n \rbrace$ now serve as coordinates of the force network.
Typically the protocol used to generate the contact network -- here we employ molecular dynamics simulations -- also produces a force network that can be used as the particular solution ${\bf f}_0$. As in a triangulation, alternative choices for ${\bf f}_0$ can be identified via simulated annealing or linear programming.

\subsubsection{Relation to the isostatic length}
Because force rearrangements are closely related to floppy modes, it is natural to expect a connection to the ``isostatic length'' $\ell^*$. The isostatic length governs several mechanical properties of packings of soft frictionless particles \cite{silbert05,wyart05,wyart05b,ellenbroek06,zeravcic09,vanhecke10} and is related to floppy modes via the ``bond cutting'' argument of Wyart and co-workers \cite{tkachenko99,wyart05,wyart05b}, which we briefly summarize. Consider an infinite or periodic packing with mean coordination number $\z = z_{\rm iso} + \Delta z$. $\z > z_{\rm iso}$ so that the packing has no floppy modes. Now imagine removing a cluster of size $\ell$ from the packing, as in Fig.~\ref{fig:arbgrain}d; in so doing one cuts ${O}(\ell^{d-1})$ ``bonds'', viz.~contacts, on the boundary. 
The rigidity of the cluster is determined by a competition between cut bonds, which remove constraints and therefore inhibit rigidity, and ``excess bonds'' (in excess of an isostatic packing) in the interior, which lend redundant rigidity to the cluster. If there are more cut bonds than excess bonds, of which there are ${O}(\Delta z\,\ell^d)$, the cluster will possess floppy modes. The marginal case occurs when the two populations balance, and this selects the isostatic length $\ell^* \sim 1/\Delta z$.

Let us now consider the Maxwell-Cremona diagram of the same packing. Its mean coordination number is below isostaticity, cf.~Eq.~(\ref{eqn:dualz}), and so the diagram contains floppy modes. We again select a cluster of grains from the packing, only now we fix the ${O}(\ell^{d-1})$ forces on its boundary. In the reciprocal tiling this isolates one portion of the tiling from the rest, see Fig.~\ref{fig:arbgrain}e. Fixing forces adds constraints, and if there are enough of them to overcome the shortfall of bonds in the interior (compared to an isostatic packing), the isolated region will be rigid.
As the shortfall of bonds is ${O}(|\Delta \tilde{z}|\,\widetilde{N}) \sim {O}(\Delta z \,\ell^d)$ and the number of fixed forces is a boundary term, the isolated region will be rigid when $\ell$ is sufficiently small. The marginal case is again $\ell^* \sim 1/\Delta z$. Recalling that floppy modes in the reciprocal tiling construct force rearrangements, we see that the isostatic length can be interpreted as the typical linear size of a grain cluster that supports just one internal force rearrangement. Indeed, it was recently shown in Ref.~\cite{mailman10} that point-to-set correlation functions in the FNE, which measure the overlap between force networks in a finite size cluster, scale with $\ell^*$.

\subsection{Additive Extensive Invariants}
\label{sec:invariants}

\begin{figure}[tbp] 
\centering
\includegraphics[clip,width=0.9\linewidth]{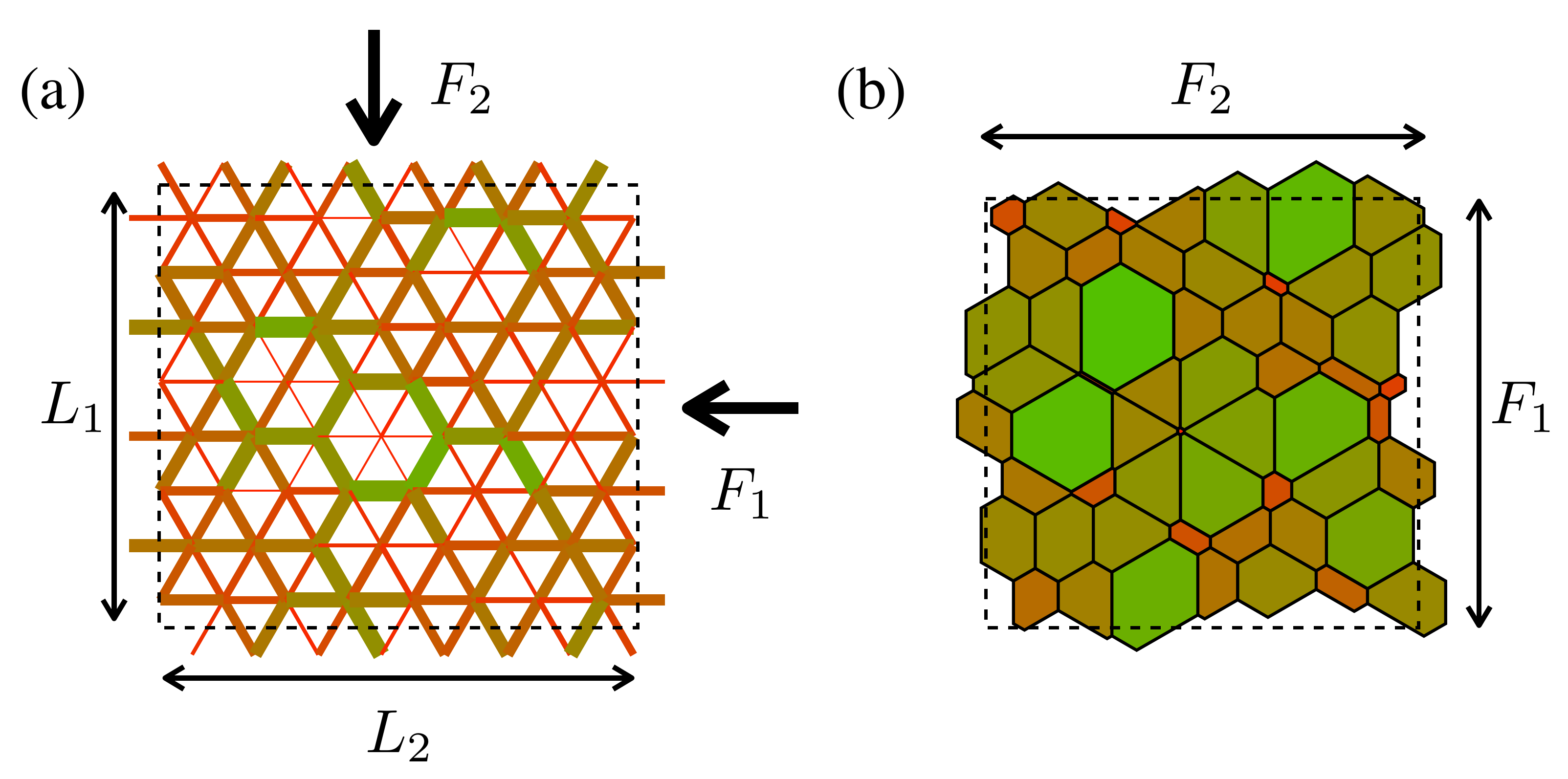}
\caption{(a) A force network from the triangular lattice and (b) its Maxwell-Cremona diagram or reciprocal tiling. }
\label{fig:periodic}
\end{figure}

We have now shown that in hyperstatic frictionless disk packings it is possible to construct a set of contact force rearrangements that transform one force balanced force network into another. The force network ensemble can then be defined as all noncohesive force networks that can be reached, starting from some initial force network ${\bf f}_0$, via these force rearrangements. We now show that the only salient feature of the initial force network is its stress tensor $\sbar$. In so doing, we will identify two additive, extensive quantities that are invariant under the Monte Carlo dynamics of the force network ensemble.

The reciprocal tiling was helpful in constructing force rearrangements; we use it again here to describe the stress tensor of a statistically homogeneous force network. Consider the periodic force network of Fig.~\ref{fig:periodic}a and its reciprocal tiling in Fig.~~\ref{fig:periodic}b. The packing has orthogonal primitive vectors ${\vec L}_1 = L_1 \hat{e}_1$ and ${\vec L}_2 = L_2 \hat{e}_2$, and any surface normal to $\hat{e}_i$  experiences a net compressive force ${\vec F}_i = \sbar {\vec L}_i$. Without loss of generality we choose $\hat{e}_1$ and $\hat{e}_2$ to align with the principal stress directions, so that
\begin{eqnarray}
\label{eqn:tilingstress}
{\bar \sigma}_{11} &=& \frac{F_1}{L_1} \nonumber \\
{\bar \sigma}_{22} &=& \frac{F_2}{L_2} \nonumber \\
{\bar \sigma}_{12} &=& 0  \,.
\end{eqnarray}
At the same time, by periodicity the tiling must have primitive vectors ${\vec F}_1$ and ${\vec F}_2$. Because the contact network is fixed in the FNE,  the ``size and shape'' of the tiling directly encodes the stress tensor via Eq.~(\ref{eqn:tilingstress}). In the following sections, the tiling's area $\cal A$ will play an important role. Note that 
\begin{eqnarray} \label{eqn:APreln}
{\cal A} &=& F_1 F_2 \nonumber \\ 
&=& ( {\rm det}\,\sbar) {\cal V} \,,
\end{eqnarray}
and is therefore extensive. From the construction of the tiling, $\cal A$ is manifestly additive: 
\begin{equation}
{\cal A} = \sum_{i=1}^N a_i \,,
\end{equation}
where $a_i$, the area of the tile corresponding to grain $i$, is
\begin{equation}
a_i = \frac{1}{2}{\hat e}_3 \cdot \sum_{j=1}^{z_i} {\vec g}_j \times {\vec g}_{j+1} \,.
\end{equation}
The sum runs over the $z_i$ contacts of grain $i$ ordered in a righthand fashion around the grain, and indices are taken modulo $z_i$. The vector ${\vec g}_{j+1} := \sum_{k=1}^j {\vec f}_{jk}$. Note ${\vec g}_1 = 0$. The unit vector ${\hat e}_3$ points out of the plane in a sense such that $a_i$ is positive when all forces are compressive.

Eq.~(\ref{eqn:tilingstress}) has important implications for the force network ensemble. To change the stress tensor or tiling area, a force rearrangement must change the primitive vectors of the reciprocal tiling's unit cell. A wheel move, disordered or not, {\em cannot} change the primitive vectors ${\vec F}_1$ and ${\vec F}_2$. To see this, consider the action of a wheel move in the tiling, as in Fig.~\ref{fig:floppy}c: the move simply shuffles area among a small number of tiles, leaving the unit cell of the tiling unchanged. Therefore the stress tensor $\sbar$, or equivalently the extensive stress $\Sbar :=  \sbar  {\cal V}$, is a topological invariant of the FNE, as is the tiling area $\cal A$. This observation applies equally to general disordered contact networks, because there force rearrangements are linear superpositions of disordered wheel moves. 

The force network ensemble therefore bears strong similarity to the microcanonical ensemble. Just as energy is an additive, extensive invariant of the dynamics in an equilibrium system, the extensive stress $\Sbar$ is an additive, extensive invariant of the Monte Carlo dynamics of the FNE. Moreover, because ``dynamics'' in the FNE are performed by a random walk in the space of force networks, force networks are sampled with equal {\em a priori} probability. This is again reminiscent of the equilibrium microcanonical ensemble. In the following, both for simplicity and to reinforce the analogy, we restrict our attention to ensembles of {\em isotropic} force networks, so that the scalar ``extensive pressure'' ${\cal P} = \frac{1}{2}{\rm Tr}\, \Sbar$ fully specifies the stress,
\begin{equation}
\Sbar = {\cal P}\,{\mathbbm 1} \,.
\end{equation}
The extensive pressure is also additive,
\begin{equation}
{\cal P} = \sum_{i=1}^N p_i \,,
\end{equation}
where
\begin{equation}
p_i = \frac{1}{2}\sum_{j=1}^N {\vec f}_{ij} \cdot {\vec r}_{ij}  \,,
\label{eqn:localpressure}
\end{equation}
and ${\vec f}_{ij}$ is the force on grain $i$ applied by grain $j$ (nonzero only when $i$ and $j$ are in contact) and ${\vec r}_{ij}$ is the vector from the center of $j$ to $i$. We pursue the analogy to equilibrium ensembles further in the following Section.

Note that in previous work, the FNE has been defined as the flatly sampled ensemble of noncohesive force networks subject to local force balance constraints {\em and} an imposed stress tensor $\sbar$ \cite{snoeijer04a,snoeijer04b}. The definition of the force network ensemble we have offered is ultimately equivalent (see Appendix A), but instead of directly imposing a conserved stress, we arrive at it naturally through consideration of local rearrangements consistent with force balance. From this perspective, the comparison to the microcanonical ensemble, and the statistical mechanics analogy developed in the following section, is more apt.

If the usual definition of the FNE is a microcanonical one, can one pass to a canonical FNE? This question was considered in detail in Ref.~\cite{tighe10b}; the answer is yes. In a canonical ensemble the extensive pressure $\cal P$ is not invariant but is allowed to fluctuate. According to the above discussion, this cannot be achieved with superpositions of local force rearrangements. A canonical FNE, therefore, must admit additional force rearrangements that change the global stress of a force network. Still restricting ourselves to isotropic stress states, just one additional rearrangement is needed; one possibility is the rescaling of all the forces in a force network by a single scalar parameter. In the reciprocal tiling this corresponds to an affine inflation or contraction of the entire tiling. In a canonical ensemble, of course, not all force networks receive equal weight; we derive the equivalent of the Boltzmann factor in the following Section.

\section{Statistics in the FNE}
\label{sec:statmech}

In the previous section we demonstrated that the force network ensemble can be built up from force rearrangements that respect the constraints of local force balance. We now turn to a study of stress statistics in the ensemble we have constructed. After writing down and maximizing entropy in the FNE, we consider pressure in the canonical ensemble, deriving the equation of state and the scaling of pressure fluctuations. We then consider the statistics of stress at the grain scale in the form of the local pressure probability distribution $P(p)$.

Because of its similarities to equilibrium ensembles, it is natural to describe the force network ensemble within a statistical mechanics framework. 
By definition -- or alternatively, the Monte Carlo dynamics described above guarantee that -- every force network in the FNE is sampled with equal {\em a priori} probability, i.e.~with a flat measure. Here we will write down an entropy, postulate that it is maximized, and show that it correctly reproduces equal {\em a priori} sampling in the microcanonical FNE. We then follow the same approach to generate a canonical FNE. This approach is in direct analogy to the standard textbook treatment of equilibrium statistical mechanics \cite{chandler}, though we spell out the steps for completeness. 

Because both $\cal P$ and $\cal A$ are additive extensive invariants of the dynamics, and because they bear a one-to-one correspondence in noncohesive isotropic force networks, see Eq.~(\ref{eqn:APreln}), one can employ either as the analog of energy in an equilibrium ensemble. This point is discussed at length in Ref.~\cite{tighe10b}. Here we will use the extensive pressure $\cal P$ for greater similarity to other approaches in the literature \cite{evesque99,kruyt02,bagi03,ngan03,goddard04,henkes07,edwards08,metzger08}.

If a force network $\bf f$ occurs with (normalized) frequency $B({\bf f})$, there is an associated entropy 
\begin{equation}
S[B] = -\int {\rm d}{\bf f} \, G({\bf f}) \, [ B({\bf f}) \ln{B({\bf f})} ] \,.
\label{eqn:entropy}
\end{equation} 
The function $G({\bf f})$ restricts the integral to ``valid'' states, and is defined to be unity when ($i$) $\bf f$ is force balanced, ($ii$) its contact forces are noncohesive and ($iii$) the force network is isotropic, i.e.~${\bar \sigma}_{11}  = {\bar \sigma}_{22}$ and ${\bar \sigma}_{12} = 0$.  $G({\bf f}) = 0$ otherwise. 
The integral may be further restricted depending on the ensemble under consideration. We postulate that $S[B]$ is maximized subject to certain constraints. 

Let us first consider the microcanonical ensemble, in which the relevant constraint is that of normalization:
\begin{equation} \label{eqn:normalization}
1 = \int_{{\cal P}({\bf f}) ={\cal P}_0} {\rm d}{\bf f}\,  
G({\bf f}) \,  B({\bf f})   \,.
\end{equation}
The integral is restricted to force networks with extensive pressure ${\cal P}_0$. The entropy $S[B]$ can be maximized subject to Eq.~(\ref{eqn:normalization}) by introducing the Lagrange multiplier $\chi$ and writing \cite{chandler}
\begin{equation}
0  = \delta \int_{{\cal P}({\bf f}) ={\cal P}_0} {\rm d}{\bf f}\, 
G({\bf f})
\biggl[-\ln{B({\bf f})} + \chi(1) \biggr] \,B({\bf f})  \,.
\label{eqn:max}
\end{equation}
Eqs.~(\ref{eqn:normalization}) and (\ref{eqn:max}) may be solved for $\chi$ and $B({\bf f})$; one finds $B({\bf f}) = 1/\Omega({\cal P}_0)$, where 
\begin{equation} \label{eqn:dos}
\Omega({\cal P}_0) = \int{\rm d}{\bf f}\, G({\bf f}) \, \delta({\cal P}({\bf f}) - {\cal P}_0) \,.
\end{equation}
Reassuringly, we find that valid force networks in the microcanonical FNE receive equal statistical weight.

An extensive discussion of the microcanical FNE can be found in Ref.~\cite{snoeijer04b}. In the present work, we will find it convenient to adopt the perspective of a canonical ensemble, in which  the extensive pressures  ${\cal P}({\bf f})$ is allowed to fluctuate; i.e. Eq.~(\ref{eqn:normalization}) is replaced by 
\begin{equation} \label{eqn:normalization2}
1 = \int {\rm d}{\bf f}\,  G({\bf f}) \,  B({\bf f})   \,.
\end{equation}
At the same time, a constraint on 
the {\em average} extensive pressure $\langle {\cal P} \rangle$ is imposed,
\begin{equation} \label{eqn:avgpressure}
\langle {\cal P} \rangle = \int {\rm d}{\bf f} \,  
G({\bf f}) \, {\cal P}({\bf f})\, B({\bf f})   \,.
\end{equation}
We demonstrate below that the microcanonical and canonical ensembles are equivalent in the usual way. 

Maximizing entropy with the Lagrange multipliers $\zeta$ and $\alpha$, now subject to the constraints of Eqs.~(\ref{eqn:normalization2}) and (\ref{eqn:avgpressure}), we have
\begin{equation}
0  = \delta \int {\rm d}{\bf f}\, G({\bf f}) 
\biggl[
 -\ln{B({\bf f})} + \zeta(1)  +  \alpha \, {\cal P}({\bf f})
\biggr] B({\bf f}) \,.
\end{equation}
The extremal distribution is
\begin{equation} \label{eqn:boltz}
B({\bf f}) = Z^{-1} \exp{(-\alpha {\cal P}({\bf f}))} \,,
\end{equation}
where $Z := e^{\zeta - 1}$.
The Lagrange multiplier $\alpha$, the inverse of which was termed {\em angoricity} by Edwards \cite{edwards08}, is chosen to satisfy
\begin{equation} \label{eqn:eos1}
\langle {\cal P} \rangle = -\frac{\partial}{\partial \alpha} \ln{Z} \,, 
\end{equation}
and the partition function $Z$ enforces normalization of $B({\bf f})$,
\begin{equation} \label{eqn:partfn}
Z(\alpha) = \int{\rm d}{\bf f}\, G({\bf f})\, \exp{(-\alpha {\cal P}({\bf f}))} \,.
\end{equation}
Note that, in perfect analogy to the equilibrium case, force networks in the canonical FNE are weighted by a ``Boltzmann factor'' $\exp{(-\alpha {\cal P})}$.

\subsection{Macroscopic quantities}
\label{sec:macro}
It is now straightforward to consider the statistics of the extensive pressure $\cal P$, including an equation of state relating the intensive parameter $\alpha$ to $\langle {\cal P} \rangle$. 

\begin{figure}[tbp] 
\centering
\includegraphics[clip,width=1.0\linewidth]{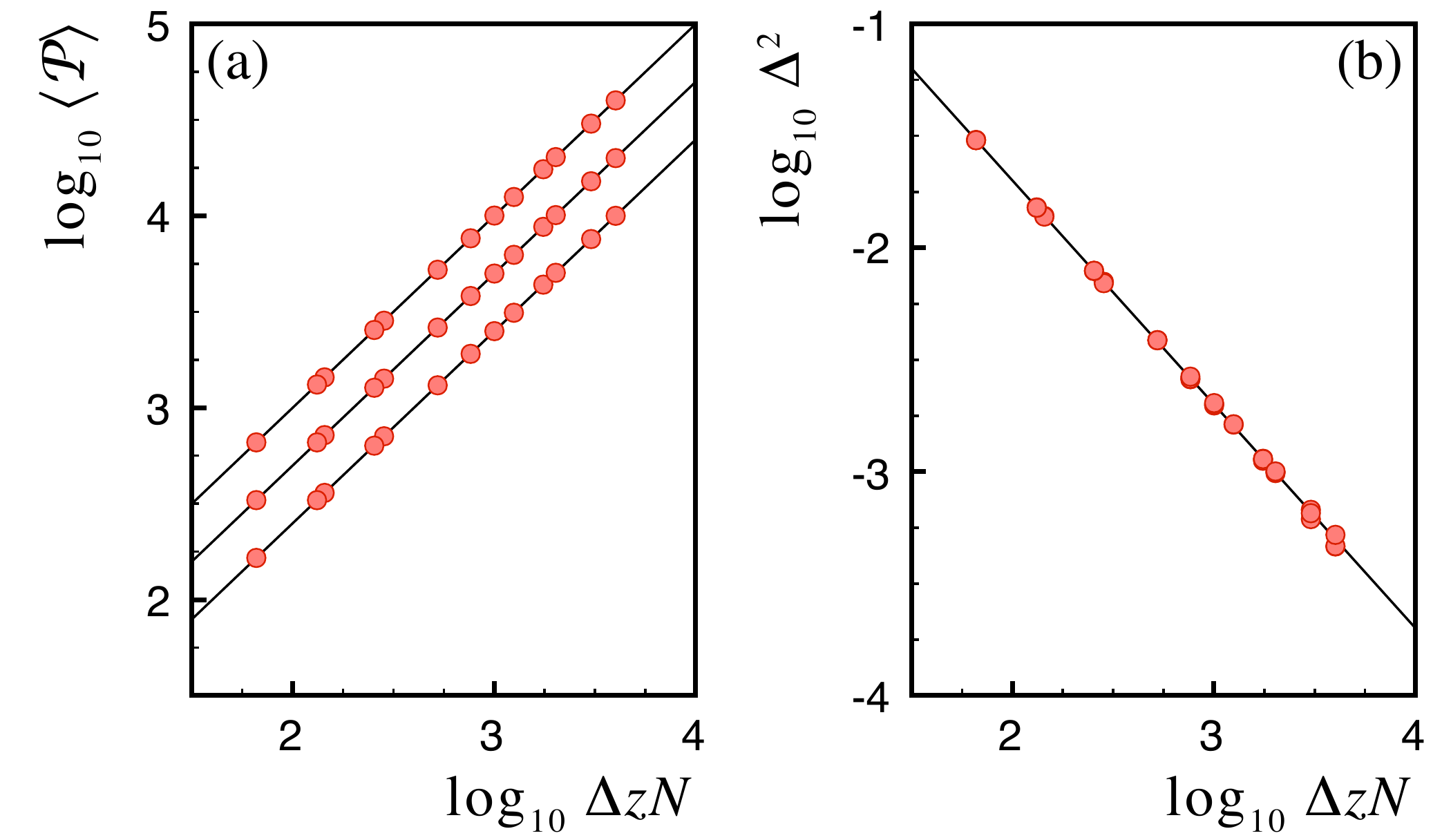}
\caption{(a) Equation of state for the FNE computed in the canonical ensemble for $\alpha = 0.05$, 0.1, and 0.2. Solid curves are Eq.~(\ref{eqn:eosFNE}). For each data point a contact network was prepared by molecular dynamic simulation of a packing of $N$ particles, with $N$ ranging from 250 to 2000; the resulting and mean coordination numbers range from $\z = 4.25$ to 6.00. Whenever necessary rattlers have been removed. 
(b) The relative fluctuations $\Delta^2 = \langle (\delta {\cal P})^2\rangle / \langle {\cal P}\rangle^2$ for the same data as in (b) collapse when plotted against $\Delta z\,N$, independent of $\alpha$, as predicted by Eq.~(\ref{eqn:relfluct}).}
\label{fig:relfluct}
\end{figure}

Our starting point is the partition function $Z$ of Eq.~(\ref{eqn:partfn}), which may be re-expressed in terms of $\Omega({\cal P})$, the density of states with extensive pressure $\cal P$:
\begin{equation} \label{eqn:partfn2}
Z = \int_0^\infty{\rm d}{\cal P}\,\Omega({\cal P})\, \exp{(-\alpha {\cal P})} \,.
\end{equation}
$\Omega({\cal P})$ has a straightforward geometric interpretation: it is the content of the space of valid force networks with extensive pressure $\cal P$. Recall that the space is a convex polytope in $N_{\rm w}$ dimensions. $\cal P$ sets  the typical ``diameter'' or linear dimension of the polytope, so that $\Omega({\cal P}) \propto {\cal P}^{N_{\rm w}}$. From Eqs.~(\ref{eqn:eos1}), (\ref{eqn:partfn2}), and (\ref{eqn:dos}), it then follows that
\begin{equation} \label{eqn:eosFNE}
\alpha \langle {\cal P} \rangle = \frac{1}{2} \Delta z  \, N + O(1)\,.
\end{equation}
Up to corrections that vanish in the thermodynamic limit, Eq.~(\ref{eqn:eosFNE}) is the equation of state of the FNE. It simply states that  $\alpha^{-1}$ selects the natural pressure scale in the ensemble. This was forseeable: having discarded the force law, the FNE does not possess an intrinsic force scale, leaving no other scale with which $\alpha^{-1}$ could compete. Eq.~(\ref{eqn:eosFNE}) is confirmed numerically in Fig.~\ref{fig:relfluct}a.

The extensive pressure fluctuations can be calculated similarly,
\begin{eqnarray}
\langle (\delta {\cal P})^2 \rangle &=& 
Z^{-1} \int_0^\infty{\rm d}{\cal P}\, \Omega({\cal P}) \, ({\cal P} - \langle {\cal P} \rangle)^2 \, \exp{(-\alpha {\cal P})} \nonumber \\
&=& N_{\rm w}/\alpha^2 \,.
\end{eqnarray}
Hence the relative fluctuations of pressure are
\begin{equation}
 \Delta^2 := \frac{ \langle (\delta {\cal P})^2 \rangle}{\langle  {\cal P} \rangle^2} =   \frac{2}{ \Delta z\,N} \,.
 \label{eqn:relfluct}
\end{equation}
Note that the pressure fluctuations are governed by $N_{\rm w}$, the dimension of the space of valid force networks.
The equivalence of canonical and microcanonical ensembles follows from the $1/N$ scaling of the relative fluctuations, which vanish in the thermodynamic limit. Note the dependence on $\Delta z$ in Eq.~(\ref{eqn:relfluct}), which ensures diverging relative fluctuations in the isostatic limit ${\Delta z \downarrow 0}$
\footnote{Of course, the limiting case itself is uninteresting in the force network ensemble: for truly isostatic packings the microcanonical FNE contains a single force network, and the canonical FNE is composed of rescalings thereof.}. Eq.~(\ref{eqn:relfluct}) is confirmed numerically in Fig.~\ref{fig:relfluct}b, which plots the pressure fluctuations in simulations of the canonical FNE. The figure shows data for a range of system sizes $N$ and coordination numbers $z$, all of which fall on the curve described by Eq.~(\ref{eqn:relfluct}). In addition, for each $(N,z)$ pair, three different values of $\alpha$ are plotted; these are difficult to see because, as predicted by Eq.~(\ref{eqn:relfluct}), the fluctuations are independent of $\alpha$.

The scaling of pressure fluctuations can also be written
$\Delta^2 = 1/\rho_{\rm w} {\cal V}$.
Therefore the pressure fluctuations are governed by the ratio of the linear system size ${\cal L} := {\cal V}^{1/d}$ to the length scale $\ell_{\rm w} :=1/\rho_{\rm w}^{1/d} \sim 1/\Delta z^{1/d}$, namely
\begin{equation}
\Delta^2  \sim \left(\frac{\ell_{\rm w}}{{\cal L}}\right)^d \,.
\label{eqn:D2}
\end{equation}
$\ell_{\rm w}$ sets the scale on which one finds fluctuations on the order of the mean pressure in periodic packings. 
We stress that $\ell_{\rm w}$ is different from, though closely related to, the isostatic length $\ell^* \sim 1/\Delta z$. The relative pressure fluctuations are not sensitive to the {\em size} of the force rearrangements, but rather to their {\em number}. Their number scales as ${\cal V}/(\ell_{\rm w})^d$ rather than ${\cal V}/(\ell^*)^d$.
This is possible because there is significant spatial overlap between the force rearrangements, i.e.~a typical contact participates in multiple  rearrangements. 

Equivalence of the microcanonical and canonical force network ensemble can be expected only for periodic systems of linear size ${\cal L} \gg \ell_{\rm w}$. 
In non-periodic systems the balance of boundary and excess bulk contacts again becomes relevant in constructing a canonical ensemble \cite{tighe10b} -- the system must at a minimum be large enough to support force rearrangements. The isostatic length $\ell^* \gg \ell_{\rm w}$ as the isostatic limit $\Delta z \downarrow 0$ is approached, hence we anticipate the stricter requirement ${\cal L} \gg \ell^*$ for canonical-microcanonical equivalence in non-periodic systems. Nevertheless, the scaling of Eq.~(\ref{eqn:D2}) must still hold for asymptotically large non-periodic systems.

\subsection{Microscopic quantities}

We now turn to the statistics of {\em local} stresses. 
One microscopic measure of the stress is the pressure $p$ on an individual grain. Although the distribution of contact forces $P(f)$ is widely studied, we will focus largely on the pressure distribution $P(p)$, which has at least two advantages over $P(f)$. First, $p$ is slightly coarse-grained with respect to $f$, making it a more realistic target for the approximate expressions we develop in later sections. Second, it will prove to be convenient that $p$, being defined on the grain scale, enters at the same scale as the constraints of local force balance.

The local pressure distribution is given by  
\begin{eqnarray}
\label{eqn:formalPp}
P_{\mu}(p) &=& [\Omega({\cal P}_0)]^{-1} \int {\rm d}{\bf f} \, 
G({\bf f})\,\delta({\cal P}({\bf f}) - {\cal P}_0) \,\delta(p_1({\bf f})-p) \nonumber \\
P_\alpha(p) &=& [Z(\alpha)]^{-1}\int {\rm d}{\bf f}  \,
G({\bf f})\,e^{-\alpha {\cal P}({\bf f})} \,\delta(p_1({\bf f})-p) 
\,,
\end{eqnarray}
within the microcanonical and canonical FNE, respectively.
Here $p_1$ is the pressure on the grain with index 1 calculated via Eq.~(\ref{eqn:localpressure}). Formally the distribution is particular to the choice of grain on which the pressure $p_1$ is assigned, unless the contact network is a Bravais lattice, though in practice little difference is found \cite{snoeijer04b}. These two distributions of Eq.~(\ref{eqn:formalPp}) converge to the same form in the thermodynamic limit, as demonstrated numerically in Ref.~\cite{tighe10b}. Therefore to study the local stress statistics one may choose to work in whichever ensemble is convenient.

Although Eq.~(\ref{eqn:formalPp}) can be solved exactly for very small systems \cite{snoeijer04b,tighe05}, one must resort to asymptotic or approximate methods in larger systems.
We first show that $P(p)$ has a power law form for asymptotically small $p$, with an exponent dictated by local topology and force balance. We then develop an expression for the full distribution $P(p)$; though the treatment is approximate, it is sufficient to capture quantitatively the pressure statistics in the frictionless triangular lattice.

\subsubsection{Statistics of small pressures}
\label{sec:smallpressures}

We now demonstrate that the local pressure distribution on a grain with $z_i$ contacts scales as $P(p_i) \sim p_i^{\nu_i}$ in the limit $p_i \downarrow 0$, where  $\nu_i = z_i - d - 1$ for frictionless spherical grains. In other words, for small pressures the statistics are governed by local topology and constraints.

The notation $\bf f$, used above, is a compact way to indicate the set of contact forces $\lbrace {\vec f}_{ij} \rbrace = \lbrace f_{ij}{\hat e}_{ij} \rbrace$, where ${\hat e}_{ij}$ indicates the unit vector aligned with the contact from $i$ to $j$. Here we employ a more explicit notation. In an enumeration of unique contact forces, each contact $(i,j)$ need appear only once because $f_{ij} = -f_{ji}$ by Newton's third law. In the following, however, we allow the product over contacts $\prod_{(i,j)}$ to include both $(i,j)$ and $(j,i)$, and explicitly write down the constraint due to Newton's third law.
\begin{widetext}
\begin{eqnarray}
\label{eqn:longPp}
P_\alpha(p_1) &=& 
Z^{-1}\int_0^\infty \left[ \prod_{(i\neq1,j)} {\rm d}f_{ij} \,\,
e^{-\alpha  r_{ij}f_{ij}/2 }\,\,
\delta\left(f_{ij} + f_{ji}\right) \right] 
\left[ \prod_{m \ge 2} \delta^{(d)}\left(\sum_{(i=m,j)} f_{ij}{\hat e}_{ij}\right) \right] J(p_1, \lbrace f_{j1} \rbrace) \,
 \nonumber \\
&{\rm where}& \nonumber \\
J(p_1, \lbrace f_{j1} \rbrace) &=&
\int_0^\infty 
\left[ \prod_{(1,j)} {\rm d}f_{1j} \,\,
e^{-\alpha  p_1 }\,\,
\delta\left(f_{1j} + f_{j1}\right)  \right]
 \delta^{(d)}\!\left(\sum_{(1,j)} f_{1j} {\hat e}_{1j}\right)
 \delta\left(\frac{1}{2}\sum_{(1,j)} r_{1j}f_{1j}  - p_1 \right)   \,.
\end{eqnarray}
\end{widetext}
The $\delta$-functions serve to impose Newton's third law and vector force balance. The extensive pressure $\cal P$ in the Boltzmann factor has been expressed in terms of the individual contact forces. Terms involving the grain indexed 1, on which the pressure $p_1$ is imposed, have been explicitly separated in the function $J(p_1, \lbrace f_{j1} \rbrace)$.

Note that the only dependence on $p_1$ enters through $J(p_1, \lbrace f_{j1} \rbrace)$. Secondly, the term $e^{-\alpha p_1} \approx 1$ for small $p_1$. Further, grain 1 ``interacts'' with other grains only via Newton's third law.  Thus, assuming interactions can be neglected in $J(p_1, \lbrace f_{j1} \rbrace)$ for asymptotically small $p_1$, the pressure distribution scales as  
\begin{eqnarray}
P_\alpha(p_1) &\sim& 
\int_0^\infty 
\left( \prod_{(1,j)} {\rm d}f_{1j}  \right) 
\delta^{(d)}\!\left(\sum_{(1,j)} f_{1j}{\hat e}_{1j}\right)
\nonumber \\
&& \,\,\,\,\,\,\,\,\,\,\,\,\,\,\,
\times \delta\left(\frac{1}{2}\sum_{(1,j)} r_{1j}f_{1j}  - p_1 \right)   \,.
\label{eqn:singlegraindos}
\end{eqnarray}

The righthand side of Eq.~(\ref{eqn:singlegraindos}) may be conceptualized in the following way. Given a grain with $z_1$ contacts, each force balanced configuration of the grain is a point in a state space in which the $z_1$ contact force magnitudes are coordinate axes. Though the number of axes suggests $z_1$ dimensions, force balanced configurations occupy a ($z_1 - d$)-dimensional volume because of the $d$ force balance constraints on the grain. Single grain force configurations with a pressure $p_1$ reside on a ($z_1 - d - 1$)-dimensional ``slice'' through this volume. The integral of Eq.~(\ref{eqn:singlegraindos}) is the content of this slice. Because the state with all contact forces being zero (hence zero pressure $p_1$) is force balanced, and because all contact forces must be non-negative, the origin is a corner of the volume of balanced force configurations. Moving out of this corner, i.e.~increasing $p_1$, the content of slices grows as $p_1^{z_1 - d - 1}$. We thus have
\begin{equation} \label{eqn:smallpnofric}
P_\alpha(p_1) \sim p_1^{z_1 - d - 1} \,,
\end{equation}
for small pressures in frictionless packings. Employing the canonical ensemble did not play an important role; repeating the calculation in the microcanonical ensemble gives the same result. This is to be expected; in the thermodynamic limit, the statistics 
of microscopic quantities should not depend on the choice of ensemble. 

We stress the implication of Eqs.~(\ref{eqn:singlegraindos}) and (\ref{eqn:smallpnofric}): for weak interactions, the scaling of the pressure probability distribution can be inferred from degree of freedom counting --- local contacts versus local force balance constraints --- and the flat measure on the ensemble.

\begin{figure}[tbp] 
\centering
\includegraphics[clip,width=0.75\linewidth]{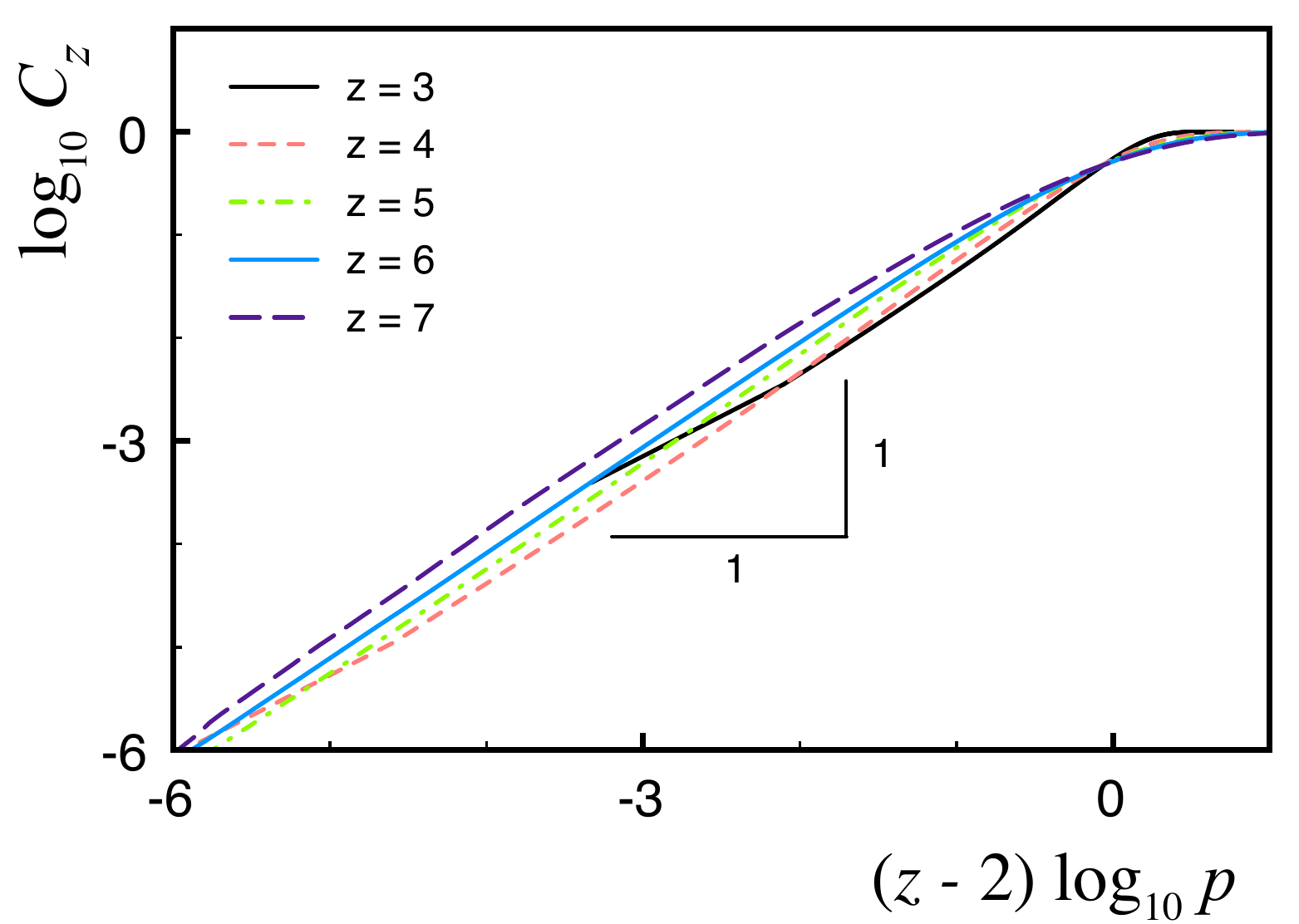}\\
\caption{ Numerically sampled cumulative distributions $C_z(p) = \int_0^p {\rm d}p^\prime P(p^\prime|z)$. Statistics are sampled in the microcanincal FNE on a periodic disordered packing in $d=2$ dimensions composed of 17 grains with 3 contacts, 233 grains with 4 contacts, 489 grains with 5 contacts, 279 grains with 6 contacts, and 6 grains with 7 contacts.
}
\label{fig:cumdist}
\end{figure}

To test these results or equivalently the reasonableness of neglecting interactions in the above calculations, we perform numerical simulations of the FNE in a disordered packing. The conditional probability distribution $P(p|z)$ is then sampled, i.e.~the probability of obtaining a pressure $p$ given that a grain has $z$ contacts. Because the resulting curves are smoother, we plot the cumulative distribution $C_z(p) := \int_0^p {\rm d}p^\prime P(p^\prime|z)$. Fig.~\ref{fig:cumdist} plots of $\log{C_z(p)}$ versus $(z-d)\log{p}$ for a frictionless system; the linearity of the curves for small $p$ confirms Eq.~(\ref{eqn:smallpnofric}). 

\subsubsection{Statistics of large pressures}
\label{sec:local}

The success of the approach in Section \ref{sec:smallpressures} indicates that in the limit of asymptotically small pressures it is possible to adopt a ``single grain picture'', i.e.~to simplify calculations by neglecting interactions with neighboring grains, and still successfully predict local pressure statistics. 
Ref.~\cite{tighe08b} demonstrated that an approximate treatment of the statistics of local pressure in a single grain picture can be surprisingly accurate, even for $p \gtrsim \langle p \rangle$. We now describe this approach.

We demonstrated above that a maximum entropy postulate correctly reproduces the appropriate equal {\em a priori} weighting of valid force networks in the microcanonical FNE, and used the same method to construct the canonical FNE. All of the results derived in this manner have direct analogs in equilibrium statistical mechanics.
We now employ the principle of maximum entropy more broadly. When all that is known about a system is that it must satisfy certain constraints, the ``best guess'' is that the system's state is described by a probability distribution that maximizes (information) entropy \cite{jaynes}. This statement reduces to the approach of Section \ref{sec:macro} if one imposes {\em all} the relevant constraints on the system, including the constraints of local force balance {\em on every grain}. Though these local constraints are conceptually unproblematic and straightforward to implement in simulation, they render expressions like Eq.~(\ref{eqn:longPp}) difficult or impossible to evaluate. Because of these technical difficulties, we will seek to make approximations. In so doing, we take a simple lesson from information theory: the more information one incorporates (in the form of constraints on the system), the more accurate one can expect the predicted pressure distribution to be.

We now perform a calculation in which entropy is maximized subject to constraints on both $\langle {\cal P} \rangle $ {\em and} $\langle {\cal A} \rangle$. We have seen that, in the presence of local force balance, these two constraints are redundant. However, we will also assume a single grain picture in which interactions with neighboring grains are not explicitly incorporated. In so doing, the mechanism by which the constraints on $\langle {\cal P} \rangle $ and $\langle {\cal A} \rangle$ are redundant is broken --- there is no tiling unless every grain is in local force balance. In this context, imposing the constraint on $\langle {\cal A} \rangle $, in addition to $\langle {\cal P} \rangle$, reintroduces some of the information that was lost by neglecting interactions. In effect, it tells the central grain something about the consequences of force balance on all the {\em other} grains in the system. The surprise is that with this one additional piece of information it is possible to describe local pressure distributions quantitatively. There is a price to be paid for this approach: because $\langle {\cal P} \rangle $ and $\langle {\cal A} \rangle$ are not truly independent constraints, the Lagrange multipliers $\alpha$ or $\gamma$ that we introduce to impose them cannot be associated with  ``true'' intensive thermodynamic parameters in the thermodynamic limit. Therefore $\alpha$ and $\gamma$ within this approximate method should not be invested with any physical significance.

We will treat the case of the frictionless triangular lattice. The main simplification comes from the ordered contact network; we saw above that the pressure distribution $P(p)$ depends on the local coordination number. In a Bravais lattice, of course, each grain has the same number of contacts; in the triangular lattice $P(p) \sim p^{z-d-1} = p^3$ for $p \downarrow 0$. We have confirmed numerically that pressure statistics in disordered systems are similar to the triangular lattice; in particular, their tails have the same qualitative form and, as shown above, they obey Eq.~(\ref{eqn:smallpnofric}). However, the disordered case requires extending the theory to account for how pressure and tiling area are distributed among subpopulations with different local coordination numbers. This is an interesting question that we leave to future work.

In light of the above discussion, we identify the single grain state with its pressure $p$ and tiling area $a$ and approximate the entropy of the system as $S = Ns$, where the single grain entropy $s$ is
\begin{equation}
\label{eqn:singlegrainent}
s[b] = - \int_0^\infty {\rm d}p \int_0^\infty {\rm d}a \, 
\upsilon(p,a)  \left[ b(p,a) \ln{b(p,a)} \right] \,,
\end{equation}
to be maximized subject to
\begin{eqnarray}
\label{eqn:singlegrain}
1 &=&\int_0^\infty {\rm d}p \int_0^\infty {\rm d}a \,
\upsilon(p,a) b(p,a) \nonumber \\
\langle p \rangle &=&\int_0^\infty {\rm d}p \int_0^\infty {\rm d}a \,
\upsilon(p,a)\, p\, b(p,a) \nonumber \\
\langle a \rangle &=&\int_0^\infty {\rm d}p \int_0^\infty {\rm d}a \,
\upsilon(p,a)\, a\, b(p,a) \,.
\end{eqnarray}
Here $\upsilon(p,a)$ is the single grain density of states with pressure $p$ and tiling area $a$. The result is
\begin{equation}
b(p,a) \propto \exp{(-\alpha p - \gamma a)} \,,
\end{equation}
and the joint distribution $P(p,a)$ is then
\begin{equation}
P(p,a) = Z^{-1} \upsilon(p,a) \exp{(-\alpha p - \gamma a)} \,.
\end{equation}
The Lagrange multipliers $Z$, $\alpha$, and $\gamma$ are determined via Eq.~(\ref{eqn:singlegrain}). 

It is convenient to factorize $\upsilon(p, a) = \omega(p)\psi(a|p)$, where $\omega(p) = \int {\rm d}a \, \upsilon(p,a)$ is the single grain density of states with pressure $p$. It is given by the righthand side of Eq.~(\ref{eqn:singlegraindos}), and therefore $\omega(p) \sim p^3$ in the frictionless triangular lattice. $\psi(a|p) = \upsilon(p,a)/\omega(p)$ is the density of single grain states with tiling area $a$, given that the grain has a pressure $p$. We will assume for now, and confirm below, that $\psi(a|p)$ is peaked at a value $a^*(p) \approx \langle a(p) \rangle$; that is, given a pressure $p$, the most likely value of the tiling area (the mode of $\psi(a|p)$) is well approximated by the mean area of tiles with pressure $p$. Under this assumption,
\begin{equation}
\int_0^\infty {\rm d}a \, \psi(a|p)  \, \exp{(-\gamma a)} 
\approx \exp{(-\gamma \langle a(p) \rangle)} \,,
\label{eqn:approx}
\end{equation}
up to a prefactor that can be absorbed in $Z$. The local pressure probability distribution is $P(p) = \int_0^\infty {\rm d}a\, P(p,a)$, which becomes
\begin{equation}
\label{eqn:Pptrinofric}
P(p) = Z^{-1} p^3 \exp{(-\alpha p - \gamma \langle a(p) \rangle)} \,.
\end{equation}
Thus the problem has been reduced to finding $\langle a(p) \rangle$.

\begin{figure}[tbp] 
\centering
\includegraphics[clip,width=0.75\linewidth]{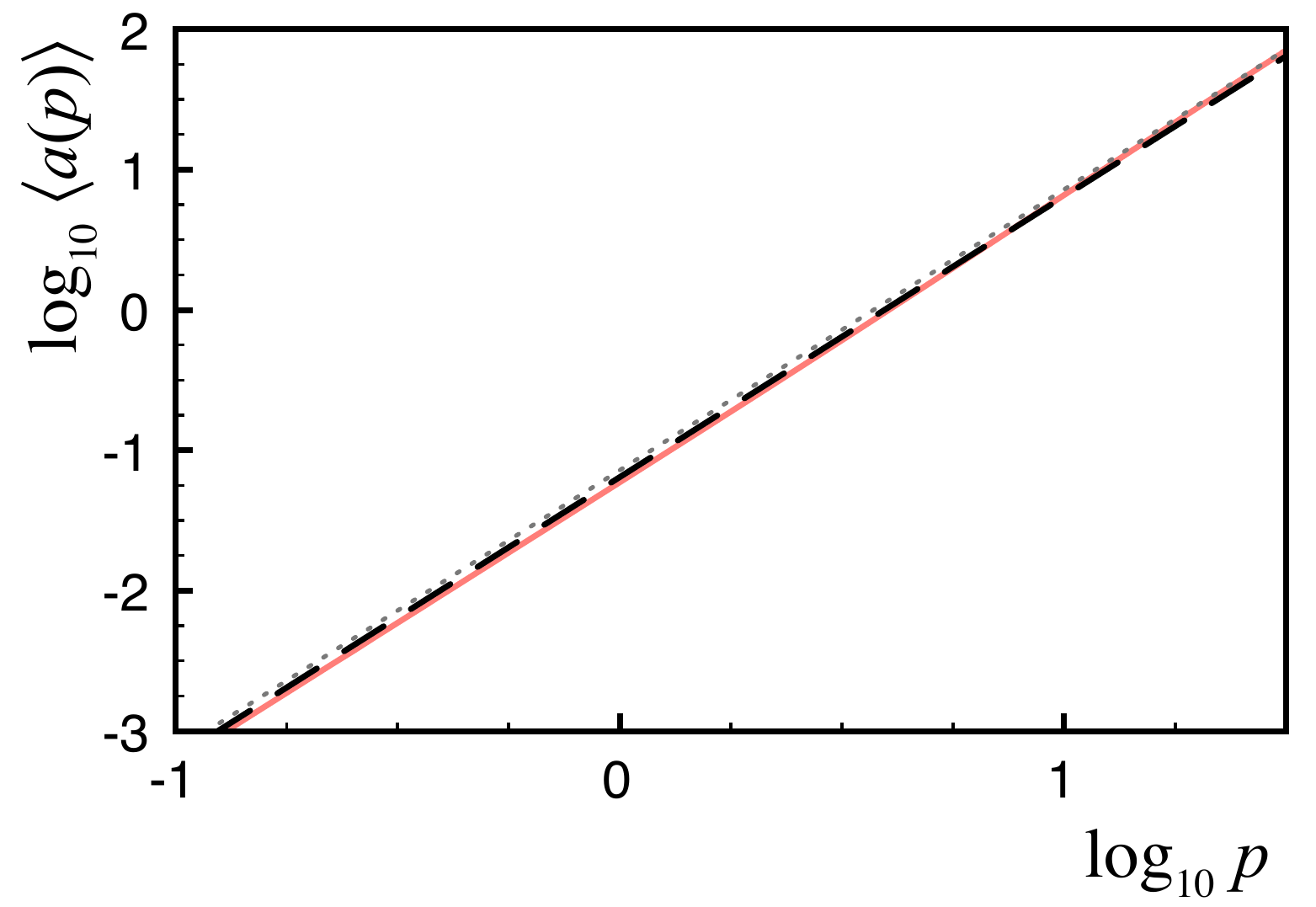}\\
\caption{ (solid curve) Numerically sampled average area $\langle a(p) \rangle$ of a tile in the triangular lattice, given that the corresponding grain has pressure $p$. (long dashed curve) Fitted quadratic function $0.893 (\langle a \rangle / \langle p \rangle)^2 p^2$. (short dashed curve) Area of a regular hexagon with perimeter $p$. }
\label{fig:areafn}
\end{figure}

Recall that in the reciprocal tiling, lengths are proportional to forces in the force network. In the triangular lattice, therefore, the pressure $p$ on a grain is directly proportional to the perimeter of the corresponding tile. Therefore, in the simplest possible scenario, one anticipates from dimensional analysis
\begin{equation}
\label{eqn:quadratic}
\langle a(p) \rangle \propto p^2 \,.
\end{equation}
In the frictionless triangular lattice, the area $a(p)$ of a tile with pressure (perimeter) $p$ is bounded by the area of a regular hexagon with the same perimeter
\begin{equation}
a(p) \le \frac{\sqrt{3}}{24}p^2 = \frac{\langle a \rangle}{\langle p \rangle^2}p^2 \,.
\label{eqn:reghex}
\end{equation}
This is plotted in Fig.~\ref{fig:areafn}, which shows that the actual behavior of $\langle a(p) \rangle$ is indeed quadratic, to good approximation, and comes close to saturating the regular hexagon bound. From a fit to the numerically sampled $\langle a(p) \rangle$ we determine 
\begin{equation}
\langle a(p) \rangle \approx 0.893 \frac{\langle a \rangle}{\langle p \rangle^2}p^2 \,.
\label{eqn:fit}
\end{equation}

\begin{figure}[tbp] 

\centering
\includegraphics[clip,width=1.0\linewidth]{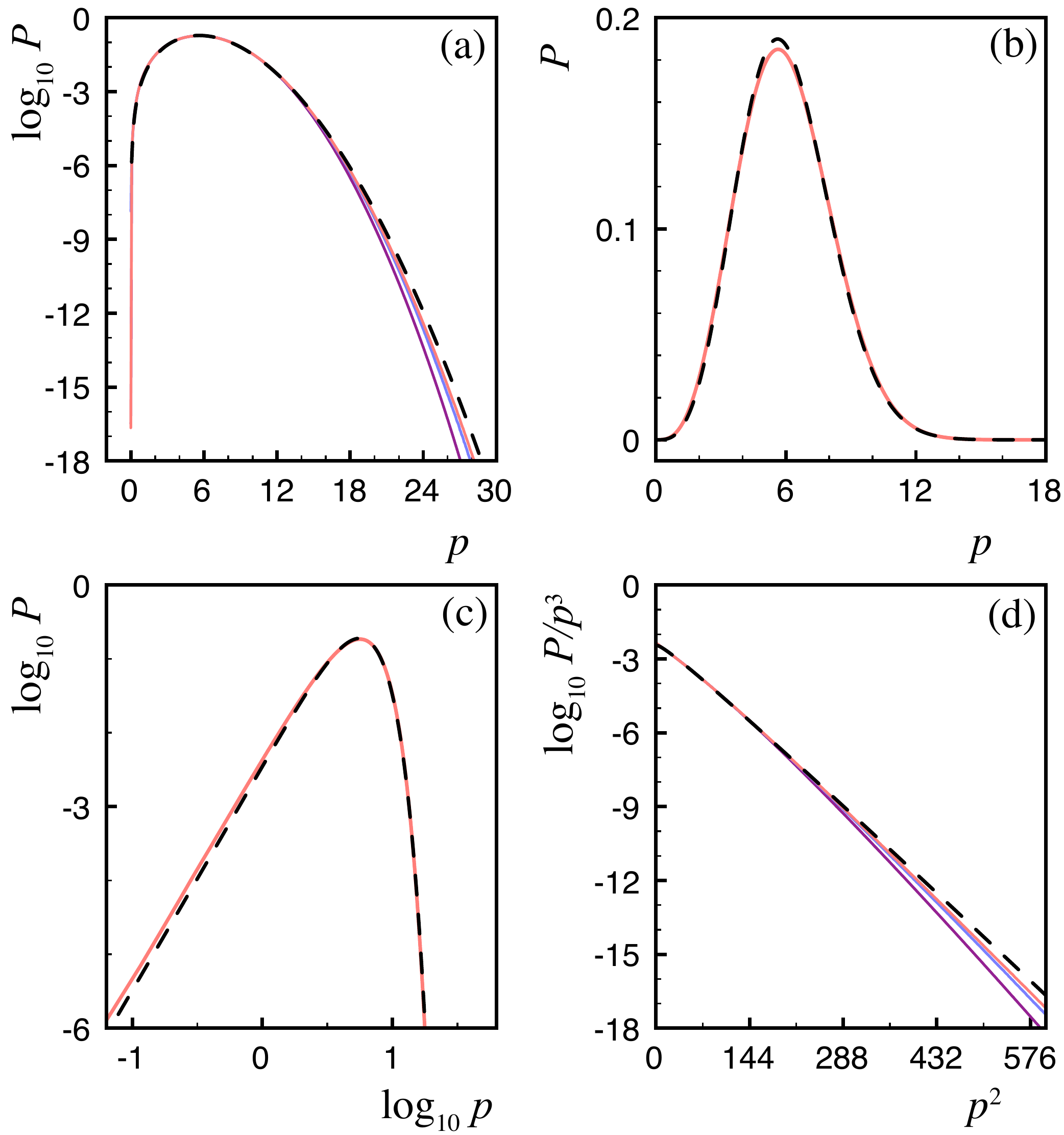}\\
\caption{(color online) Numerics (solid curves) and Eq.~(\ref{eqn:Pptrinofric}) (dashed curves) for the local pressure probability distribution $P(p)$ in the frictionless triangular lattice of $N = 1840$ grains. The same data is shown in (a) semi-log, (b) linear-linear and (c) log-log plots, as well as (d) $\log{(P/p^3)}$ versus $p^2$. To demonstrate finite size effects, (a) and (d) also contain data for $N = 460$ and $N = 115$.}
\label{fig:Pptrinofric}
\end{figure}

The pressure distribution $P(p)$, determined using 
Eq.~(\ref{eqn:fit}), is plotted in Fig.~\ref{fig:Pptrinofric}. The agreement with the numerically sampled distribution is excellent. Numerical distributions are computed using umbrella sampling, which allows us to determine the tail of $P(p)$ extremely accurately; see Appendix C for a description of the method. The prediction of Eq.~(\ref{eqn:Pptrinofric}) captures the cubic growth at small $p$ (Fig.~\ref{fig:Pptrinofric}b), the peak near $p \approx \langle p \rangle = 6$ (Fig.~\ref{fig:Pptrinofric}c), and the Gaussian tail. The latter feature is best seen in Fig.~\ref{fig:Pptrinofric}d, which plots $\log{P(p)/p^3}$ versus $p^2$. The numerically sampled distributions approach a line with slope $-1$. Finite size systems fall off somewhat faster than a Gaussian, but deviations from a Gaussian tail decrease with increasing system size $N$. 
Note that, had the mean area of a tile $\langle a \rangle$ not been imposed, we would have recovered a distribution with $\gamma = 0$, i.e.~an exponential tail, in clear disagreement with numerics. Thus the extra information provided by the tiling constraint has allowed us to capture the Gaussian tail of $P(p)$.

\subsection{Boundaries} \label{sec:boundary}
\begin{figure}[tbp] 
\centering
\includegraphics[clip,width=1\linewidth]{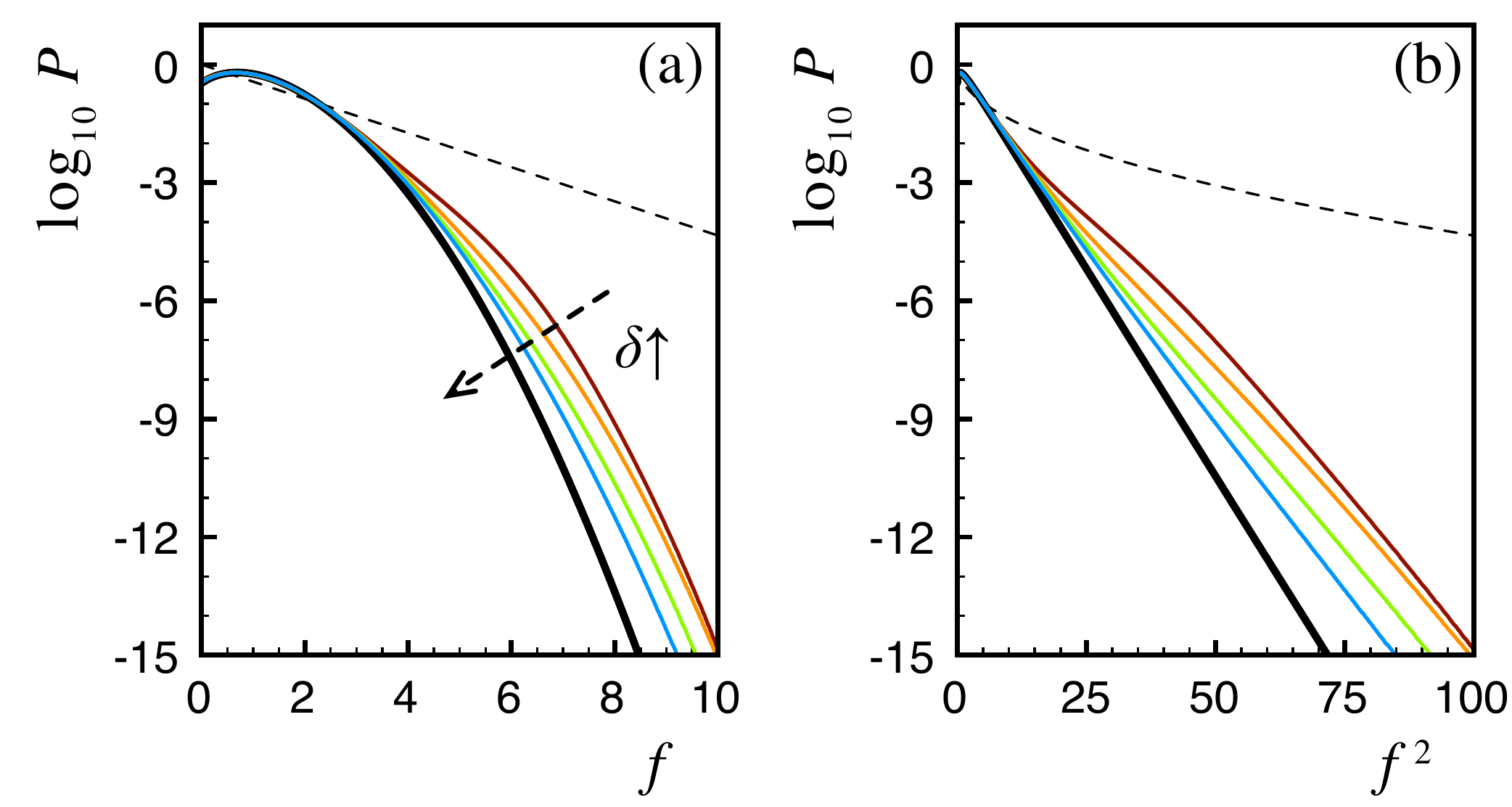}\\
\caption{ (a) Probability distribution $P(f)$ of forces in the bulk of the frictionless triangular lattice with a boundary. Boundary forces are selected from an exponential distribution (dashed curve). Distributions are plotted for all contacts more than $\delta=2$, 5, 7, and 10 layers distant from the boundary (thin curves). The distributions approach the form of $P(f)$ in the bulk of the periodic frictionless lattice subject to a flat measure (thick curve). (b) Data from (a) plotted versus $f^2$.}
\label{fig:boundary}
\end{figure}

All the preceding discussion has been concerned with the statistics of stresses in the bulk of a granular packing. Here we consider the influence of boundaries.

We introduce a boundary along a triangular lattice direction and subject it to a compressive force $F$. In the orthogonal direction the packing remains periodic  and is subject to a compressive stress such that stress tensor $\sbar$ is isotropic. In an experiment a packing is prepared by loading its boundary, rather than controlling the stress in the bulk directly. It is therefore reasonable to assign equal statistical weight to every microscopic configuration of the {\em boundary forces} consistent with the macroscopic load, i.e.~the compressive force $F$. This is a subtle departure from the standard prescription in the microcanonical FNE, in which each force network consistent with a macroscopic stress carries equal statistical weight --- all the networks of the flatly sampled FNE remain, but the measure in the space of force networks is no longer flat. A similar measure was proposed in Ref.~\cite{blumenfeld07}, though there the ensemble was restricted to isostatic states. In the thermodynamic limit one should expect, and we confirm below, that the distinction between a flat measure on all force configurations and a flat measure on the boundary configurations is not significant. We now show that, although the statistics of local stress in the bulk are indeed unaffected, the distinction {\em is} important for the statistics at the boundary.

We consider a frictionless triangular lattice with boundaries normal to the $y$-direction and periodic boundary conditions in the $x$-direction. A total normal force $\sum_i (f_{\rm b})_i = F$ is imposed on the boundaries. For equally-weighted boundary force configurations $\lbrace (f_{\rm b})_i \rbrace$, the distribution of boundary forces $P(f_{\rm b})$ is exponential, $P(f_{\rm b}) = \langle f_{\rm b} \rangle^{-1} \exp{(-f_{\rm b}/\langle f_{\rm b} \rangle)}$. Here $\langle f_{\rm b} \rangle = F/N_{\rm b}$ and $N_{\rm b}$ is the number of boundary contacts. Despite the exponential boundary force distribution, we empirically observe that the tail of the force distribution a short distance from the boundary remains Gaussian, as in the case of a flat measure on the space of all force networks. Fig.~\ref{fig:boundary} plots the force distribution on grains at least $\delta$ layers from the boundary for $\delta = 2$, 5, 7, and 10. The force distribution $P(f)$ quickly approaches its form in the bulk of a periodic packing subject to a flat measure (thick black curve). Although the evolution of force statistics with depth is itself an interesting question, we leave its analysis to future work.

The flat boundary force measure is interesting because it automatically provides an exponential force distribution on the boundary, in good agreement with experimental boundary force measurements \cite{liu95,mueth98,lovoll99}. Within the force network ensemble we find that, even with an exponential distribution of boundary forces, the tail of the force distribution in the bulk remains Gaussian. Thus, whether or not the flat boundary measure is realistic, it provides a straightforward example of a system in which the boundary and bulk force distributions are qualitatively different. This suggests that one must be cautious when inferring even qualitative features of the bulk distribution, such as the form of the tail, from experimental measurements on the boundary.

\subsection{Spatial correlations} \label{sec:correlations}
\begin{figure}[tbp] 
\centering
\includegraphics[clip,width=1.0\linewidth]{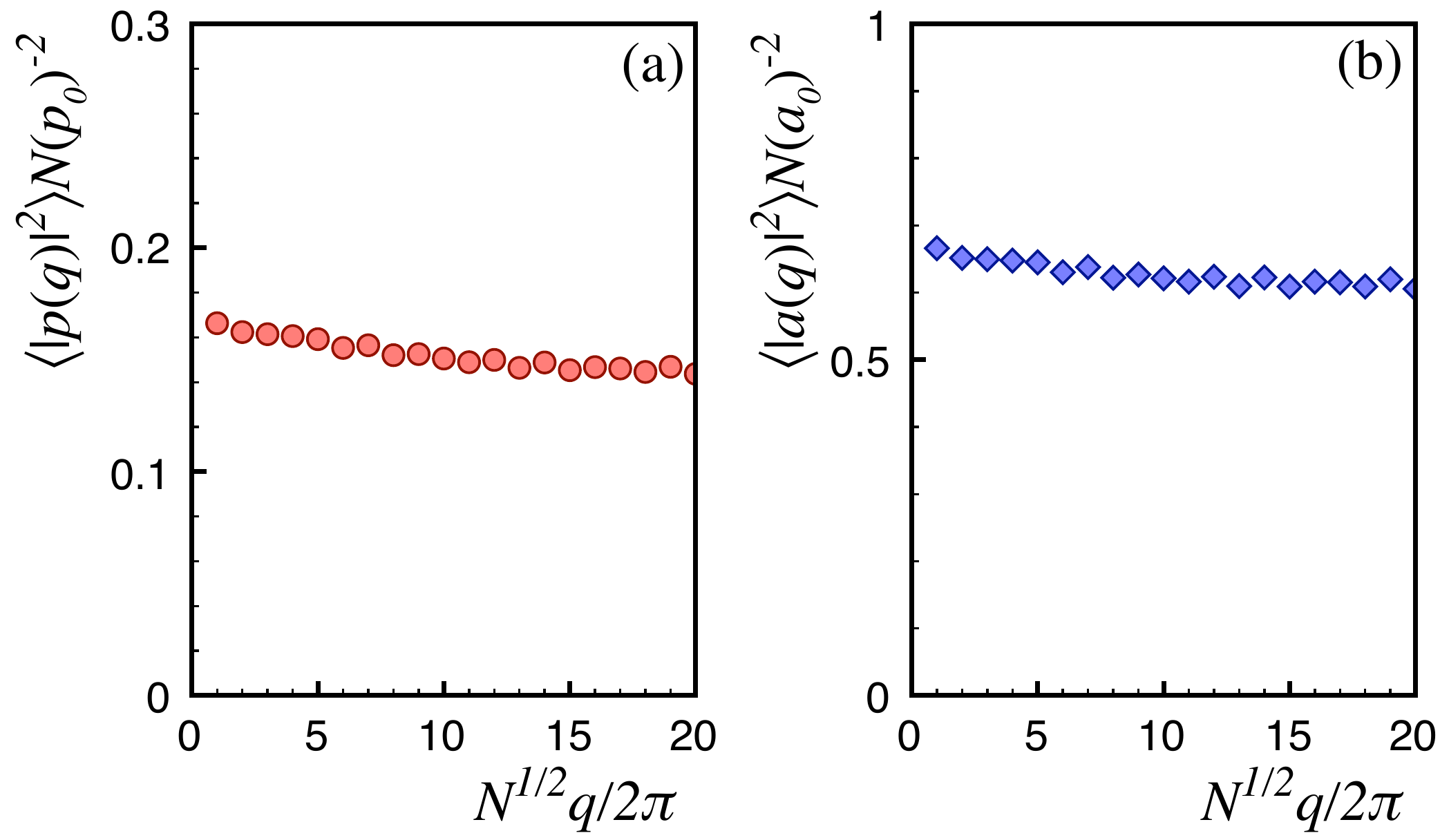}\\
\caption{ The structure factors (a) $\langle |p({\vec q})|^2 \rangle$ and (b) $\langle |a({\vec q})|^2 \rangle$ in the triangular lattice with $N=1840$ grains. The wavevector $\vec q$ is parallel to a reciprocal lattice basis vector. The rescaling of the $y$-axis is suggested by Eq.~(\ref{eqn:loosepositivity}). For uncorrelated variables the structure factor is flat. }
\label{fig:structurefactor}
\end{figure}

A single-grain picture, which was employed above to describe $P(p)$ in the triangular lattice, cannot be expected to succeed in the presence of strong interactions, i.e.~strong spatial correlations. We therefore seek now to characterize the correlations in the triangular lattice.
Correlations may be conveniently characterized by the structure factors
$\langle |p({\vec q})|^2 \rangle$ and $\langle |a({\vec q})|^2 \rangle$,
where $\varphi({\vec q}) = N^{-1}\sum_{\vec r} \varphi({\vec r}) e^{i{\vec q}\cdot{\vec r}}$ is the spatial Fourier transform of the position-dependent function $\varphi({\vec r})$. A flat structure factor indicates the absence of spatial order. Fig.~\ref{fig:structurefactor} shows that the pressure and area structure factors for the triangular lattice is nearly flat, confirming that correlations are indeed weak.

The results of Fig.~\ref{fig:structurefactor} may be partially motivated by considering the nontensile constraint. Each contact can sustain a normal force $f$ which must be compressive; under our sign convention, this corresponds to a positivity constraint $f \ge 0$. A weaker constraint, which follows from positivity of the forces, is positivity of the pressure $p$ on each grain. The converse is not true; pressure positivity does not guarantee force positivity. Nevertheless, arguments derived from pressure positivity are useful for developing intuition.

A stress state composed of the mean pressure $\langle p \rangle = p_0$ modulated by a single oscillatory mode $p_{\vec r} = p_0 + p_{\vec q}\cos{({\vec q}\cdot {\vec r})}$ must obey $|p_{\vec q}| \le p_0$. Noting that the resulting constraint on fluctuations is independent of $q$, we now assume that there is a typical scale $\tilde p$ for each $|p_{{\vec q} \neq 0}|$ and make a random phase approximation. Requiring $\langle (p_{\vec q} - p_0)^2\rangle \lesssim p_0^2$ then gives
\begin{equation} \label{eqn:loosepositivity}
\frac{\tilde{p}}{p_0} \lesssim \frac{1}{\sqrt{N}}\,.
\end{equation}
A similar argument can be made for the area fluctuations. The ordinate axes in Fig.~\ref{fig:structurefactor} have been rescaled to show that Eq.~(\ref{eqn:loosepositivity}) holds.

Though spatial correlations are weak, they do have some influence on the local stress statistics. We now consider further the form of the area function $\langle a(p) \rangle$, which must be determined in order to predict $P(p)$ via Eq.~(\ref{eqn:Pptrinofric}).
We argued above that $\langle a(p) \rangle \propto p^2$ is to be expected both on dimensional grounds and because $\langle a(p) \rangle$ is bounded by the area of the regular hexagon with perimeter $p$.
We now show that, in a system truly devoid of interactions, one indeed has purely quadratic scaling of $\Abarp$, as in Eq.~(\ref{eqn:quadratic}). We also show that there are in fact small corrections to quadratic scaling, which can therefore be attributed to spatial correlations.

Recalling that forces in a frictionless sytem are directed along contact normals, it is clear that the restriction to noncohesive forces also requires $a \ge 0$. For grain 1 with bond vectors $\lbrace {\vec r}_{1j} \rbrace$ there is also a maximum possible area $a_\mathrm{max}(p; \lbrace {\vec r}_{1j} \rbrace)$. We have already noted that, in the triangular lattice, this maximum area corresponds to the regular hexagon with perimeter $p$. 
When interactions are neglected, each single grain microstate has equal {\em a priori} probability, independent of $p$ and $a$.
Moreover, whether the packing is ordered or disordered, scaling all the forces from a particular microstate of a grain by a scalar $\lambda>0$ produces a new force balanced state such that $p \rightarrow \lambda p$ and $a \rightarrow \lambda^2 a$. Therefore, given the set of balanced states available for a particular $p>0$, we may produce all the states for $p^\prime$ simply by scaling each microstate by $\lambda = p^\prime/p$. 
The implication is that the conditional density of states $\psi(a|p)$ satisfies 
\begin{equation}
\psi(a|p) = \lambda^2 \psi(\lambda^2 a| \lambda p) \,.
\end{equation}
Substituting this relation in $\Abarp = \int_0^\infty {\rm d}a\, a \psi(a|p)$, one finds $\Abarp = c p^2 $
for some constant $c$. For packings under compressive stress $\langle a \rangle >0$, and therefore $c>0$. 

\begin{figure}
\includegraphics[width=0.5\linewidth]{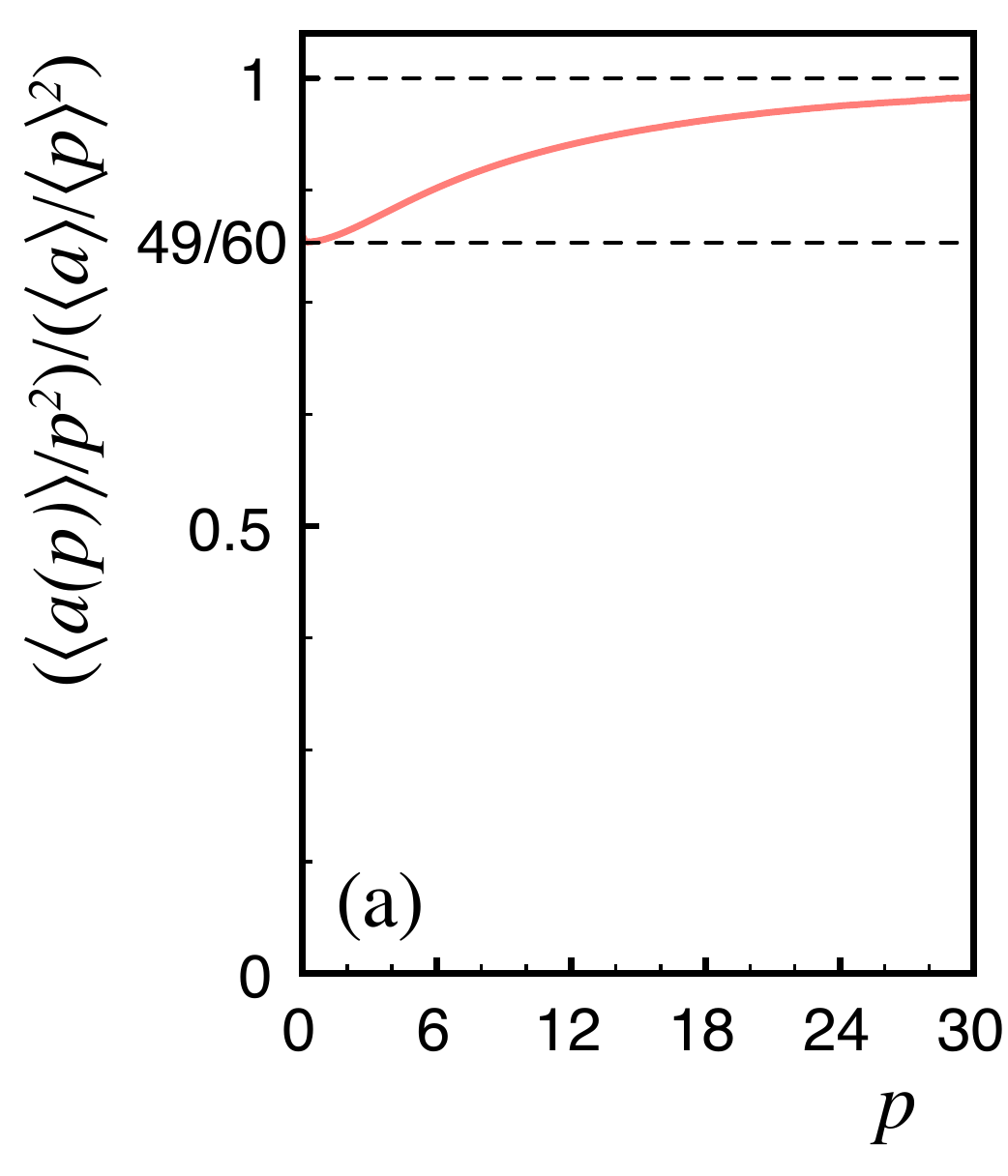}
\hspace{-0.04\linewidth}
\includegraphics[width=0.5\linewidth]{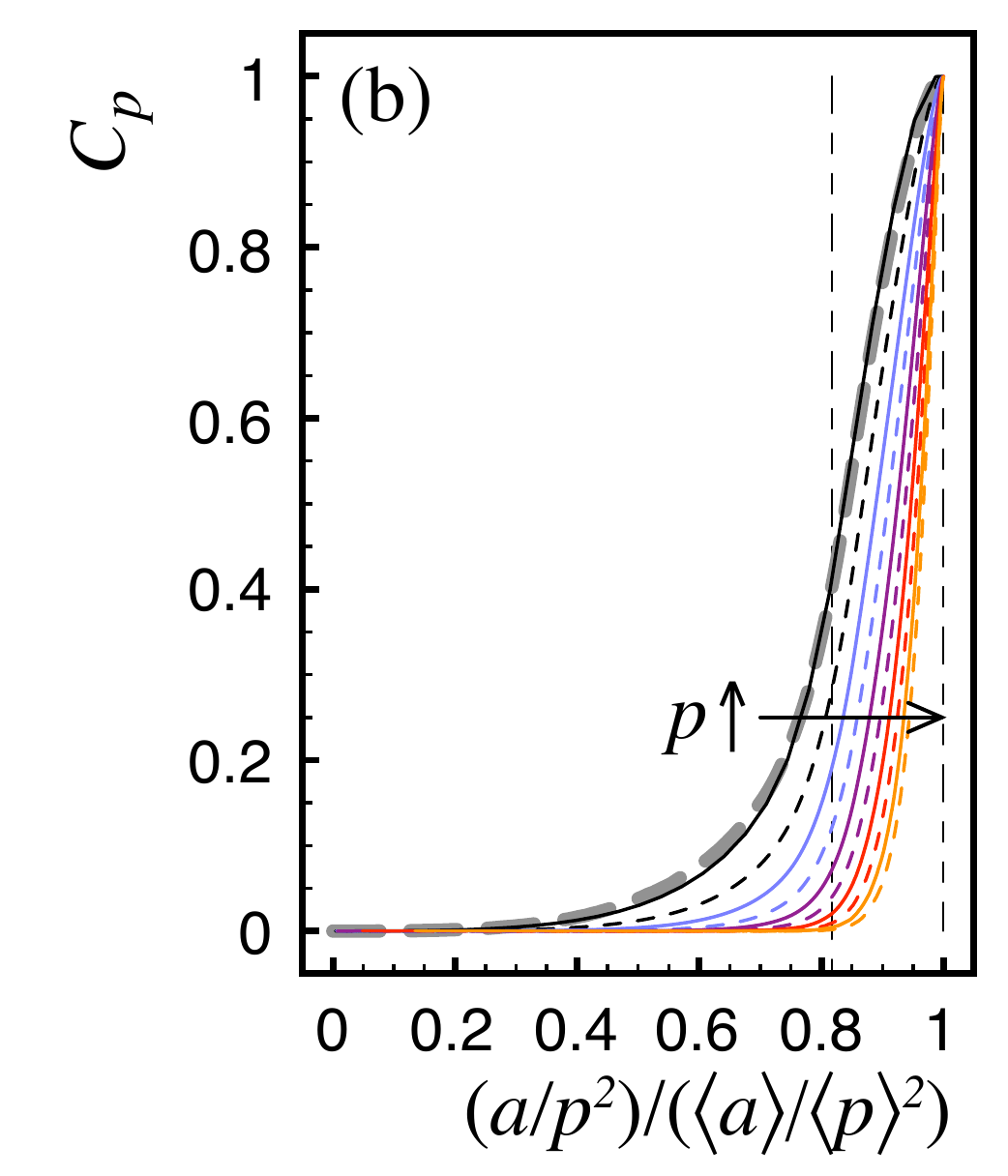}
\caption{
(a) Rescaled average area $p^{-2}\langle a(p) \rangle = p^{-2}\int_0^\infty da\,a \Psi(a|p)$ of a tile from a grain with pressure $p$. Dividing by $p^{2}$  makes it apparent that $\langle a(p) \rangle$ interpolates between two quadratic scalings. At small $p$ the coefficient approaches that predicted by a calculation neglecting spatial correlations. For large $p$ the coefficient approaches the upper bound given by a regular hexagon of perimeter $p$.
(b) Statistics of tiles with an area $a$ given that the tile has pressure ($\sim\,$perimeter) $p$. For smoother curves the cumulative distribution $C_p(a) = \int_0^a da^\prime P(a^\prime|p)$ is plotted. (thin curves) Cumulative distributions for $p = 2, 4, 6, \ldots,20$, obtained using umbrella sampling. (thick dashed curve) $C_p(a)$ of a single-grain state absent correlations with neighboring grains. Curve determined by numerically sampling all {\em single grain} force balanced states with a fixed pressure. Dashed vertical lines correspond to the asymptotes in (a).
}
\label{fig:cumulative}
\end{figure}

Because $\Abarp$ must be quadratic in systems without interactions, deviations from quadratic scaling are evidence of interactions.
To show that there are (weak) interactions in the triangular lattice, we calculate the coefficient $c$ directly assuming their absence. Namely,
\begin{eqnarray} \label{eqn:coeff}
\langle a(p) \rangle &=&
\eta^{-1}\int_0^\infty {\rm d}^6 f \, a(f_1,f_2,f_3,f_4,f_5,f_6) \times  \nonumber \\
&& \delta \left(\sum_{i=1}^6 {\vec f}_i \right) \, \delta \left(\sum_{i=1}^6 f_i - p \right) \\ 
& = & \frac{49}{60} \frac{\langle a \rangle}{\pbar^2} \, p^2
 \approx  0.82  \frac{\langle a \rangle}{\pbar^2} \, p^2 \,
 \nonumber
\end{eqnarray}
where $a(f_1,f_2,f_3,f_4,f_5,f_6)$ is the area of a tile given the grain's six forces and $\eta = \int {\rm d}^6 f \, \delta \left(\sum_{i=1}^6 {\vec f}_i \right) \, \delta \left(\sum_{i=1}^6 f_i - p \right)$. 
Replotting the data of Fig.~\ref{fig:areafn} by dividing out a quadratic scaling in $p$, as in Fig.~\ref{fig:cumulative}a, reveals that $\Abarp$ in fact interpolates between two quadratic scalings. For asymptotically small $p$, the area function is given by Eq.~(\ref{eqn:coeff}), while for asymptotically large $p$ it obeys Eq.~(\ref{eqn:reghex}) in equality. Therefore Eq.~(\ref{eqn:fit}) should be interpreted as an effective scaling that compromises between these two quadratic scalings.

Similar behavior can also be seen in the conditional probability distribution $P(a|p)$. For smoother curves, Fig.~\ref{fig:cumulative}b plots the integrated function $C_p(a) = \int_0^a da^\prime\, P(a^\prime|p)$.  If there were no spatial correlations in the system, plotting $C_p(a)$ against $a/p^2$ would collapse $C_p$ to a master curve independent of $p$. This master curve is the single grain density of states $\psi(a|p)$ in the absence of correlations, which can be obtained directly from Monte Carlo simulation of single grain force balanced states.
Deviations from the master curve indicate that there are some correlations in the system, consistent with the structure factors in Fig.~\ref{fig:structurefactor}. As expected from consideration of Fig.~\ref{fig:cumulative}a, for small $p$ the cumulative distribution is in good agreement with the master curve. For asymptotically large $p$ the function $C_p(a)$ approaches a step function near the upper bound corresponding to regular polygons, and $C_p(a)$ rises steeply over the whole range in $p$. This validates the approximation that $\int da\, \psi(a|p)e^{-\gamma a} \approx  e^{-\gamma \Abarp}$, i.e.~Eq.~(\ref{eqn:approx}).

\subsection{Higher dimensions}
\label{sec:3d}

\begin{figure}[tbp] 
\centering
\includegraphics[clip,width=0.75\linewidth]{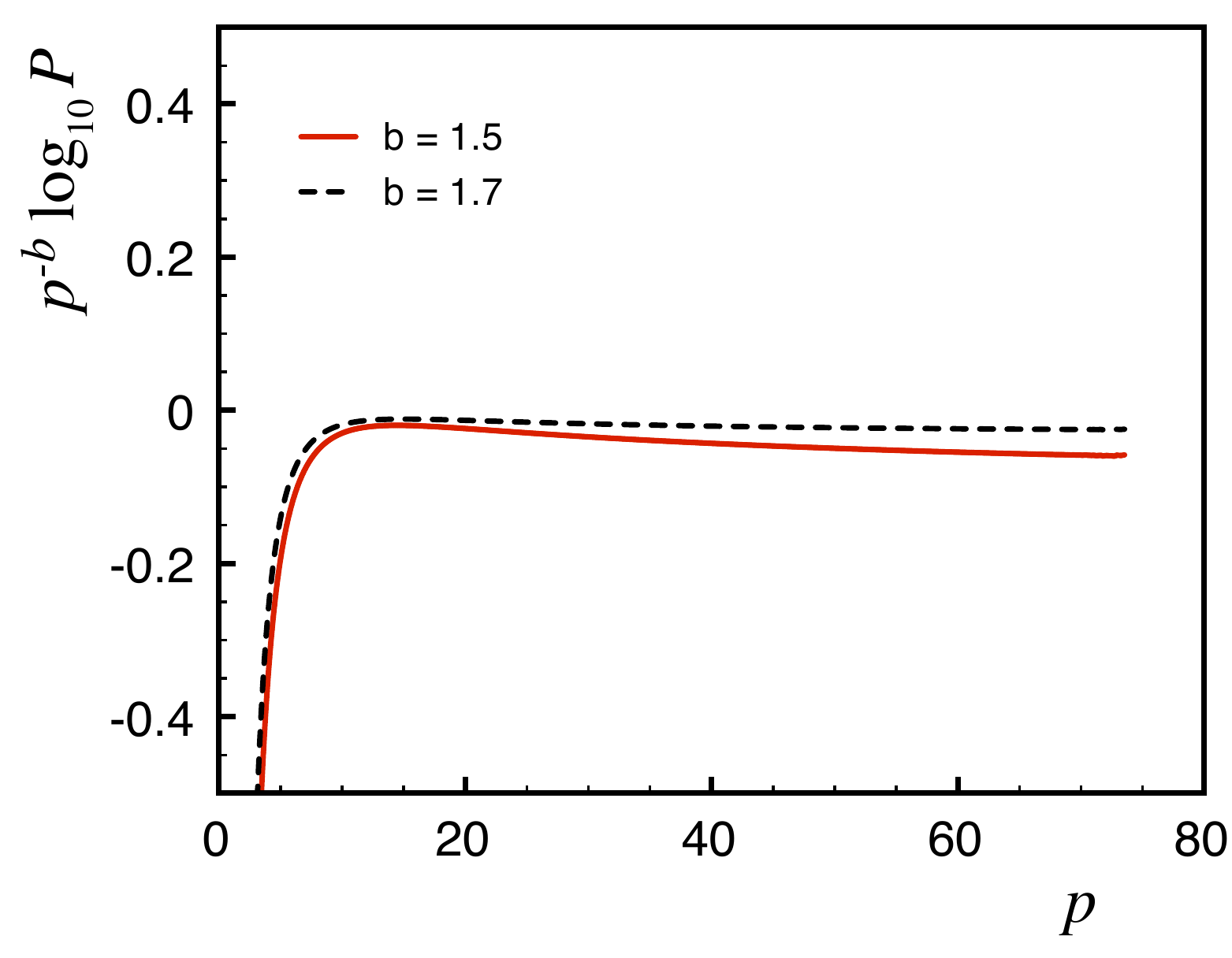}
\caption{(color online) (a-c)  Pressure distribution computed in the frictionless fcc lattice. For a distribution decaying asymptotically as $P(p) \sim \exp{(-p^b)}$, a plot of $p^{-b}\log_{10}{P(p)}$ will approach a flat line. Here $b = 1.5$ and 1.7 (see legend). }
\label{fig:Pp3d}
\end{figure}

The form of the tail of $P(f)$ in the force network ensemble depends on dimensionality $d$, as reported in Refs.~\cite{vaneerd07,tighe10}. There the authors find $P(f)$ decays asymptotically as $\exp{(-f^{b(d)})}$, where $b = 2(1)$, $1.7(1)$ and $1.4(1)$ for $d=2$, $3$, and $4$, respectively. The arguments presented above, in particular the tiling area $\cal A$, are specific to two dimensions. We now briefly consider dimensions $d \ge 3$.

A theorem due to Minkowski states that for every set of $d$-dimensional vectors $\lbrace {\vec f}_{ij} \rbrace$,  $j = 1 \ldots z_i$, $z_i>d$, such that $\sum_j {\vec f}_{ij} = 0$, there exists a unique polyhedron in $d$ dimensions with the property that the unit vectors $\lbrace {\hat f}_{ij} \rbrace$ give the outward normals to the $z_i$ faces of the polyhedron and the scalar magnitudes $\lbrace f_{ij} \rbrace$ are  the $(d-1)$-dimensional areas of the corresponding faces \cite{minkowski}. The vector contact forces on each grain satisfy $\sum_j {\vec f}_{ij} = 0$ absent body forces, regardless of dimension, and $z_i > d$ if grain $i$ is not a rattler, so with each grain we can always associate a unique ``Minkowski polyhedron''. Clearly for $d=2$ the corresponding polyhedra are the tiles of the Maxwell-Cremona diagram. For $d \ge 3$ the polyhedra do not tesselate space; it is straightforward to construct counterexamples in the frictionless simple cubic lattice \cite{herminghausprivate}. 
Thus the method whereby we demonstrated conservation of tiling area in $d=2$ cannot be generalized to $d \ge 3$. We stress, however, that tesselation is merely a means to an end; absence of a tesselation does not imply that the sum of polyhedra volumes cannot be conserved. 

Let us consider the consequences of a conjecture that in higher dimensions there exists a conserved quantity ${\cal A}^{(d)} = \sum_i a_i^{(d)}$, where $a_i^{(d)}$ is the $d$-dimensional volume of the Minkowski polyhedron of the contact forces on grain $i$. Then it follows by the entropy maximization arguments of Section \ref{sec:local} that $P(p) \sim \exp{(-p^{b(d)})}$ as $f \rightarrow \infty$, where $b(d) = d/(d-1)$, in reasonable agreement with the results of Ref.~\cite{vaneerd07}. This is tested in Fig.~\ref{fig:Pp3d}, which plots $p^{-b} \log_{10}{P(p)}$ from the frictionless fcc lattice for both $b=1.5$ and $b=1.7$. Although the plot is flatter over a broader range in $p$ for $b=1.7$, the difference between the two curves is small and a slow approach to a tail consistent with $b=1.5$ cannot be ruled out.

\section{Discussion and Outlook}
\label{sec:outlook}

We have demonstrated that ensembles of hyperstatic force networks subject to constraints of mechanical equilibrium can be described within a statistical mechanics framework. For a given contact geometry, the ensemble can be explored via force rearrangements that respect local force balance. These rearrangements are closely related to floppy modes, and their number is in proportion to the distance to isostaticity, $\Delta z$.  On a macroscopic scale, the number of force rearrangements governs the fluctuations of stress, and therefore pressure fluctuations in the FNE diverge in the isostatic limit. This divergence is characterized by a length scale $\ell_{\rm w} \sim \Delta z^{-1/d}$.

Local stress statistics can also be explored within the force network ensemble, and we have extracted considerable details regarding the distribution of grain scale pressures, $P(p)$. In the limit of small pressures the distribution scales as a power law $P(p) \sim p^{z-d-1}$, with an exponent that reflects the local connectivity of the network and local force balance constraints. In the limit of large pressures the distribution displays a Gaussian tail in two dimensions; this may be understood as a consequence of the invariance of the reciprocal tiling area $\cal A$, which is quadratic in the forces.

The force network ensemble is a minimal model; it is useful to the degree that it captures features of static granular matter in simulations and experiments, but also insofar as it points out necessary ingredients of more realistic theories. In this sense it is complementary to recent experimental attempts to identify relevant state variables in static \cite{pugnaloni10} and driven \cite{lechenault10} athermal systems. One striking feature of the FNE is the Gaussian decay (in two dimensions) of the probability density of local pressures $p$ or forces $f$. This contrasts with early measurements of boundary forces \cite{liu95,mueth98,lovoll99}, though we saw in Section \ref{sec:boundary} that statistics at the boundary need not correspond to bulk statistics. Early theoretical efforts such as the q-model also predicted exponential tails but only incorporated {\em scalar} force balance \cite{coppersmith96}; we have seen that Gaussian tails are intimately connected to the reciprocal tiling, which requires vector force balance. We consider the form of the tail of $P(f)$ in real granular systems an interesting and open question; Ref.~\cite{tighe10} summarizes recent theoretical, numerical, and experimental work on the matter.

The FNE is easily extended to frictional packings, and this is an obvious avenue of future research. The ensemble may also be used to study departures from equal {\em a priori} sampling of states, which is permitted -- even likely -- in non-equilibrium ensembles.

\begin{acknowledgments}
We have benefited from helpful conversations with a number of people, including A.R.T.~van Eerd, W.G.~Ellenbroek, M.~van Hecke, S.~Henkes, S.~Herminghaus, J.H.~Snoeijer, J.E.S.~Socolar, and Z.~Zeravcic. BPT acknowledges support from the Netherlands Organization for Scientific Research (NWO).
\end{acknowledgments}

\appendix

\section{Degree of freedom counting}

Conventionally, the microcanonical FNE on a given frictionless sphere packing has been defined as the set of force networks satisfying the matrix equation 
\begin{equation}
{\bf A}\,{\bf f} = {\bf b} \,,
\label{eqn:fne}
\end{equation}
and a set of inequalities \cite{snoeijer04a,snoeijer04b,vaneerd07,vaneerd09}. The inequalities restrict ${\bf f}$ to noncohesive states in which each $f_{ij} >0$.
${\bf A}$ is a $\z N/2 \times (dN+d(d+1)/2)$ matrix. Its first $dN$ rows encode force balance on the given contact network, i.e. $\sum_j A_{1j} [{\bf f}]_j = F_{1x}$, the $x$-component of the net force on grain 1, $\sum_j A_{2j} [{\bf f}]_j = F_{1y}$, and so on. The final $d(d+1)/2$ rows of $\bf A$ are chosen so as to return each of the unique stress tensor components of ${\bf f}$, namely ${\bar \sigma}_{\alpha \beta} = (1/2{\cal V})\sum_{ij} [\vec{r}_{ij}]_\alpha [{\vec f}_{ij}]_\beta$. Correspondingly, the first $dN$ elements of $\bf b$ are zero: each grain experiences no net force (here we assume for simplicity a periodic packing absent body forces). The final entries of $\bf b$ contain the desired values of the stress tensor components. The matrix $\bf A$ is underdetermined, and therefore solutions to Eq.~(\ref{eqn:fne}) are of the form
\begin{equation}
{\bf f} = {\bf f}_0 +  \sum_{n=1}^{N_{\bf A}} c_n {\bf f}_n^{\rm A} \,.
\label{eqn:solns}
\end{equation}
${\bf f}_0$ is a particular solution to Eq.~(\ref{eqn:fne}) and the $\lbrace {\bf f}^{\rm A}_n \rbrace$ are null vectors of $\bf A$. $N_{\bf A}$ is the nullity of $\bf A$. The similarity of Eq.~(\ref{eqn:solns}) to Eq.~(\ref{eqn:genfne}) is obvious, and in fact what we have done in Section \ref{sec:rearrangements} is to construct the null vectors of ${\bf A}$ ``by hand'', without writing down $\bf A$ itself, simply by considering the constraints of local force balance -- the force rearrangements $\lbrace \delta {\bf f}^{\rm ext}_n \rbrace$ preserve local force balance and the stress tensor components, and therefore are null vectors of $\bf A$. Here we show that $N_{\bf A} = N_{\rm w}$, i.e.~that the number of null vectors of $\bf A$ does indeed equal the number of force rearrangements we identified.

In a periodic static packing, zero net force on the entire system follows directly from periodicity. Hence $\sum_i {\vec F}_i = \sum_i \sum_j {\vec f}_{ij} = 0$, or
\begin{equation}
\sum_{i=1}^{N-1} \sum_j {\vec f}_{ij} + \sum_j {\vec f}_{Nj} = 0 \,.
\end{equation}
Therefore it suffices to impose force balance on $N-1$ of the grains; force balance on the final follows ``for free''.
This fact can be used to count the packing's (force) degrees of freedom. A periodic triangulation with $N$ vertices has $3N$ edges, and hence supports $3N$ compressive forces. There are $2(N-1)$ force balance constraints in a two-dimensional system, hence $3N - 2(N-1) = N+2$ degrees of freedom in the forces. Further constraining the three components of the stress tensor (a microcanonical FNE) leaves $N_{\bf A}=N-1$ degrees of freedom. If the contact network is not a triangulation there is one less degree of freedom for each deleted edge, namely $N_{\bf A}=N-N_{\rm d}-1 = (1/2)\Delta z\,N - 1$. (Recall $\z = 6$ in a triangulation.) Therefore $N_{\bf A} = N_{\rm w}$.  Though we have not proven linear independence of the disordered wheel moves, we have checked for, and always found, linear independence in numerically generated triangulations.

$N_{\rm w} = (1/2)\Delta z\,N - 1$, rather than $(1/2)\Delta z \,N$, because  there is a null direction associated with the $N$ wheel moves in a triangulation. We now construct that null direction. In a triangulation each vertex $i$ has an associated local pressure difference $\delta p_i$, expressed with respect to the pressure on grain $i$ in reference state ${\bf f}_0$. It is straightforward to see that the $\lbrace \delta p_i \rbrace$ can be formulated as a linear superposition of the disordered wheel move weights, $\delta p_i = \sum_j L_{ij} w_j$. (In fact the square matrix $L_{ij}$ is a discrete representation of the Laplacian operator.) There are $N$ $\delta p$'s, but one is redundant because in the microcanonical FNE $\sum_i \delta p_i = 0$ for every force network in the ensemble (see Section \ref{sec:invariants}). Therefore the $\lbrace w_j \rbrace$ satisfy the sum rule $\sum_j l_j\,w_j = 0$, where $l_j = \sum_i L_{ij}$, and one of the $w$'s can be expressed as a linear combination of the other $N-1$.

\section{Rearrangements in frictional packings} \label{sec:fricmoves}

In frictional disk packings it is again possible to construct force rearrangements from localized objects in the Delaunay triangulation. Although these objects are no longer floppy modes in the reciprocal tiling, the tiling is again helpful in identifying them. Once the local rearrangements have been identified, they can be combined to produce spatially extended force rearrangements in the packing in a manner completely analagous to Section \ref{sec:extendedmoves}. 

As in the frictionless case, we will identify a set of motions of the tiling vertices that constitute local rearrangements of the forces. Because the motions are in the tiling, the rearrangements are guaranteed to respect force balance. In the frictionless case, the additional concern was ensuring that, after rearrangement, all forces remained normal to their contacts. In the frictional case this restriction is lifted, but we must now be sure that the rearrangement respects local torque balance.

\begin{figure}[tbp] 
\centering
\includegraphics[clip,width=0.75\linewidth]{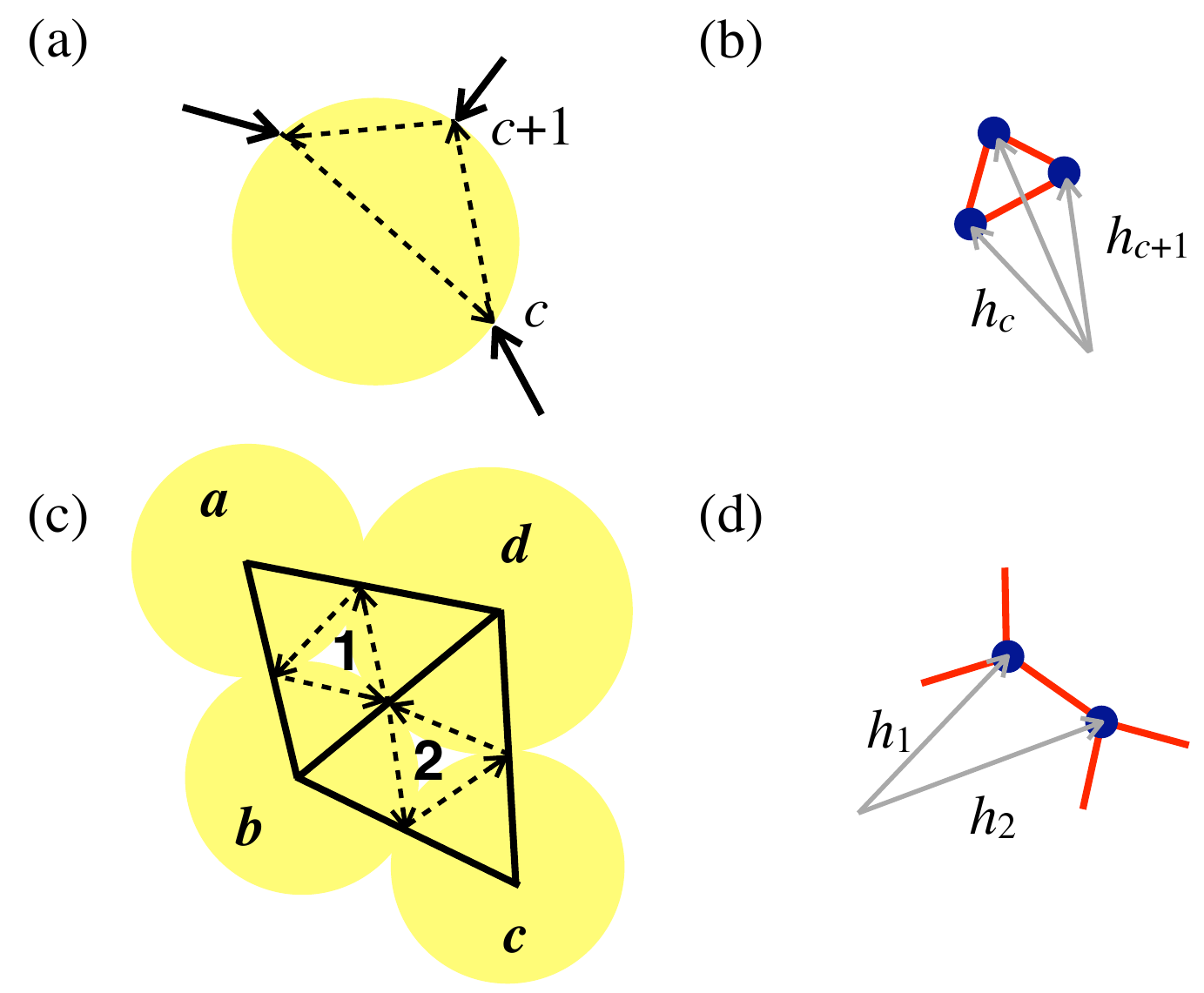}\\
\caption{ (a) Indexing of the $z_i$ contacts around grain $i$ subject to frictional contact forces (black arrows).  (b) Reciprocal tile of the grain in (a). Coordinates of the tiling vertices are labeled $\lbrace {\vec h}_c \rbrace$. Vertex ${\vec h}_c$ is connects edges $c$ and $c+1$ (modulo $z_i$). (c) The building block of a frictional force rearrangement involves four grains $a$, $b$, $c$ and $d$, and two voids $1$ and $2$.  (d) Coordinates of the tiling vertices ${\vec h}_1$ and ${\vec h}_2$ corresponding to voids $1$ and $2$ of (c). }
\label{fig:localfric}
\end{figure}

Consider a disk as in Fig.~\ref{fig:localfric}a with contacts located at $\lbrace {\vec R}_{c} \rbrace$ and forces $\lbrace {\vec f}_{c} \rbrace$ acting on those contacts. Without loss of generality we place the origin at the disk's center. Torque balance requires $ \sum_c {\vec R}_c \times {\vec f}_c = 0$. Following Ref.~\cite{blumenfeld02}, it is convenient to rewrite this expression in the following way. Label the $z$ contacts of the grain $c = 1 \ldots z$ in a righthand fashion (Fig.~\ref{fig:localfric}a). Also label the $z$ vertices of the grain's reciprocal tile $c = 1 \ldots z$ such that contact $c$ connects vertices $c$ and $c+1$ (modulo $z$). We first construct the set of vectors $\lbrace \vec{\varrho}_c \rbrace$ depicted in Fig.~\ref{fig:localfric}a (dashed arrows), where $\vec{\varrho}_{c} := {\vec R}_{c+1} -  {\vec R}_c$. Secondly, note that the contact force ${\vec f}_c$ is the difference of two vertex coordinates ${\vec h}$ in the tiling rotated by $\pi/2$ (Fig.~\ref{fig:localfric}b), i.e.~${\vec f}_c = -({\vec h}^\perp_{c+1} - {\vec h}^\perp_{c})$. Using these definitions, the torque balance constraint on the grain is
\begin{equation}
\sum_c {\vec \varrho}_c \times {\vec h}^\perp_c = 0 \,.
\label{eqn:tb}
\end{equation}

Eq.~(\ref{eqn:tb}) is useful in constructing a localized force rearrangement. Consider the configuration of four grains labeled $a$ to $d$ in Fig.~\ref{fig:localfric}c. There are six relevant $\vec \varrho$ vectors. Adjusting our previous notation, we label, e.g., ${\vec \varrho}_{1a}$ the vector around void 1 contained in grain $a$. Similarly, there is a tiling vertex associated with each of the voids; we label them ${\vec h}_1$ and ${\vec h}_2$. We wish to identify a {\em change} in each, $\delta{\vec h}_1$ and $\delta{\vec h}_2$, that respects torque balance. Such a change will enter the torque balance constraint, Eq.~(\ref{eqn:tb}), on each grain in the figure. There are then four constraints,
\begin{eqnarray}
0 &=& {\vec \varrho}_{1a} \times \delta{\vec h}_1  \nonumber \\
0 &=& {\vec \varrho}_{1b} \times \delta{\vec h}_1 + {\vec \varrho}_{2b} \times \delta{\vec h}_2  \nonumber \\
0 &=& {\vec \varrho}_{2c} \times \delta{\vec h}_1  \nonumber \\
0 &=& {\vec \varrho}_{1d} \times \delta{\vec h}_1 + {\vec \varrho}_{2d} \times \delta{\vec h}_2   \,.
\label{eqn:tbconstraints}
\end{eqnarray}
Hence there are four unknowns, the components of $\delta{\vec h}_1$ and $\delta{\vec h}_2$, and four constraints. As in the frictionless case, however, there is a degeneracy in the constraints. Consider the sum of the four constraints of Eq.~(\ref{eqn:tbconstraints}). One finds
\begin{equation}
0 = 
\left({\vec \varrho}_{1a}+{\vec \varrho}_{1b}+{\vec \varrho}_{1d} \right) \times \delta{\vec h}_1 + 
\left({\vec \varrho}_{2b}+{\vec \varrho}_{2c}+{\vec \varrho}_{2d} \right) \times \delta{\vec h}_2 \,. 
\end{equation}
This equality holds independently of $\delta{\vec h}_1$ and $\delta{\vec h}_2$ because the terms in parentheses are sums around closed loops. Therefore Eq.~(\ref{eqn:tbconstraints}) represents only three independent constraints, and the resulting one parameter family of motions is a localized force rearrangement in a frictional packing.

Every edge in a Delaunay triangulation is shared by a pair of triangles, hence we can identify one such rearrangement for each of the $\Delta z \, N$ edges in the triangulation. For general disk packings, these frictional disordered wheel moves can be constructed in the packing's Delaunay triangulation and then combined to form extended force rearrangements that change neither the normal nor the tangential force components on deleted edges; this is done in a manner entirely analogous to the treatment in Section \ref{sec:rearrangements}.

\begin{figure}[tbp] 
\centering
\includegraphics[clip,width=0.6\linewidth]{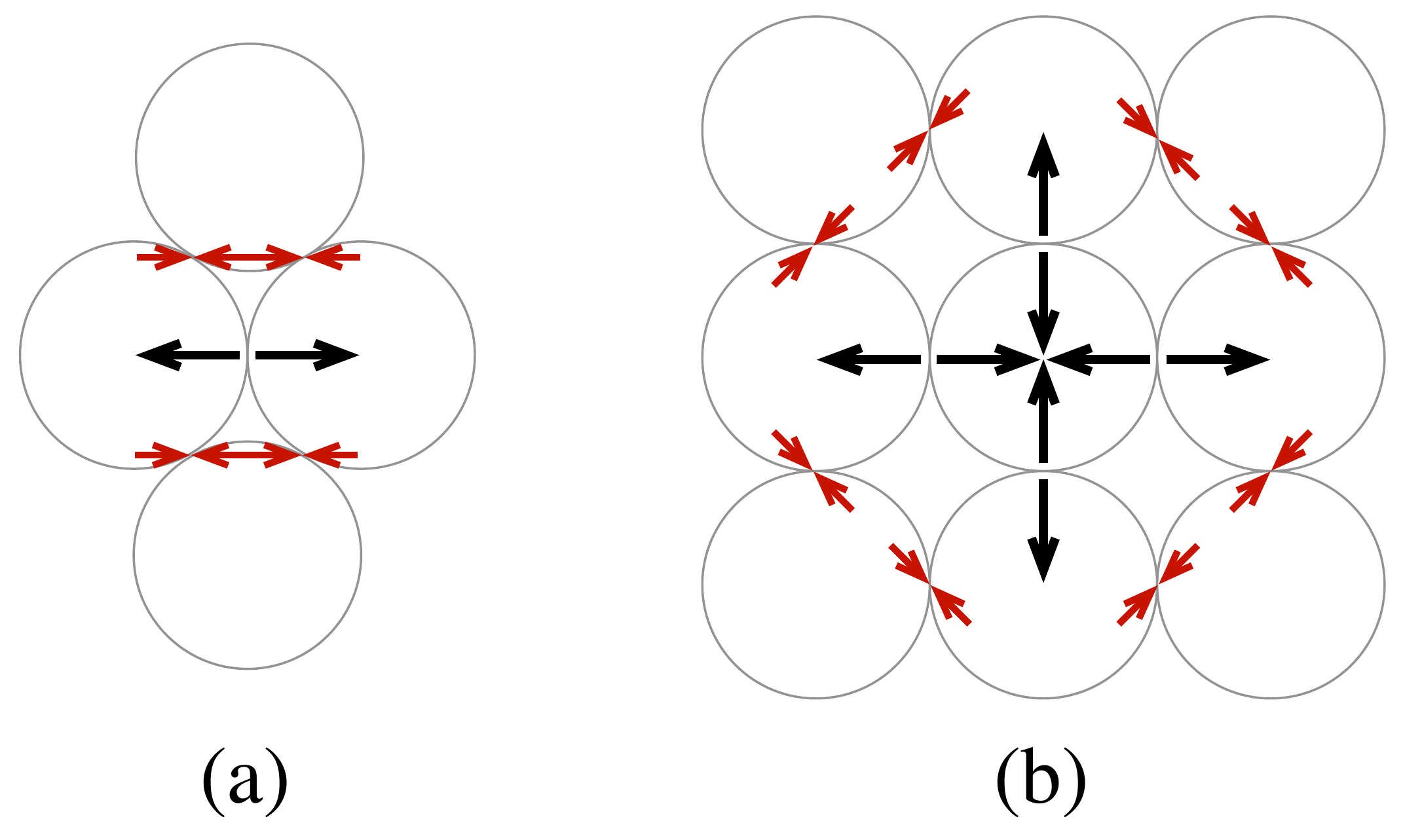}\\
\caption{ The counterparts to wheel moves in (a) the frictional triangular lattice and (b) the frictional square lattice. Arrows indicate the vector changes in force at each contact.}
\label{fig:orderedfric}
\end{figure}

In Fig.~\ref{fig:orderedfric} we depict the equivalent of wheel moves in two ordered frictional contacts networks, the triangular and square lattices, which were employed in Ref.~\cite{tighe08b}. The move in the triangular lattice corresponds directly to the situation in Fig.~\ref{fig:localfric}, while the move in the square lattice is a superposition of multiple elementary building blocks.

\section{Umbrella sampling}

\begin{figure}
\includegraphics[width=0.9\linewidth]{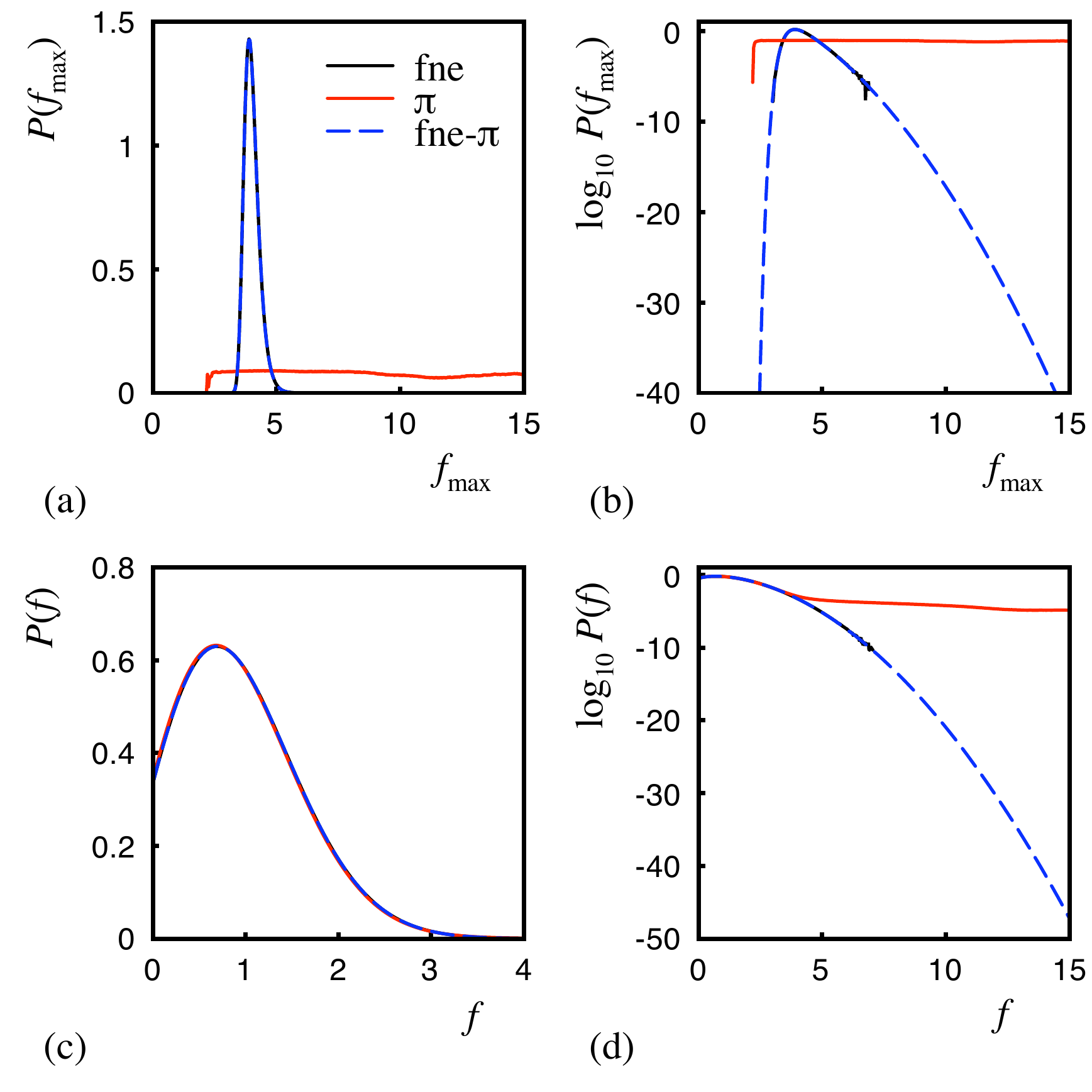}
\caption{(a,b) Probability distributions of the maximum force $f_{\rm
max}$ in a force network ${\bf f}$ and (c,d) contact force
distributions $P(f)$ for a frictionless triangular lattice
($N=1840$). Three different situations are considered: ({\em i}) the
force network ensemble without umbrella sampling (fne), ({\em ii}) the
ensemble $\pi$ in which $W(f_{\rm max})$ is chosen such that
$P_\pi(f_{\rm max})$ is approximately flat ($\pi$), and ({\em iii}) the
ensemble $\pi$ in which ensemble averages are reweighted to the force
network ensemble using Eq. \ref{eq:mappingtofne} (fne-$\pi$). In all
cases, the results for umbrella sampling (fne-$\pi$) are identical to
those computed without umbrella sampling (fne). All forces are
normalized such that $\left\langle f\right\rangle = 1$. Contact forces
larger than $5\left\langle f \right\rangle$ are hardly sampled in the
force network ensemble unless umbrella sampling is applied.}
\label{fig:um1}
\end{figure}

\begin{figure}
\includegraphics[width=0.75\linewidth]{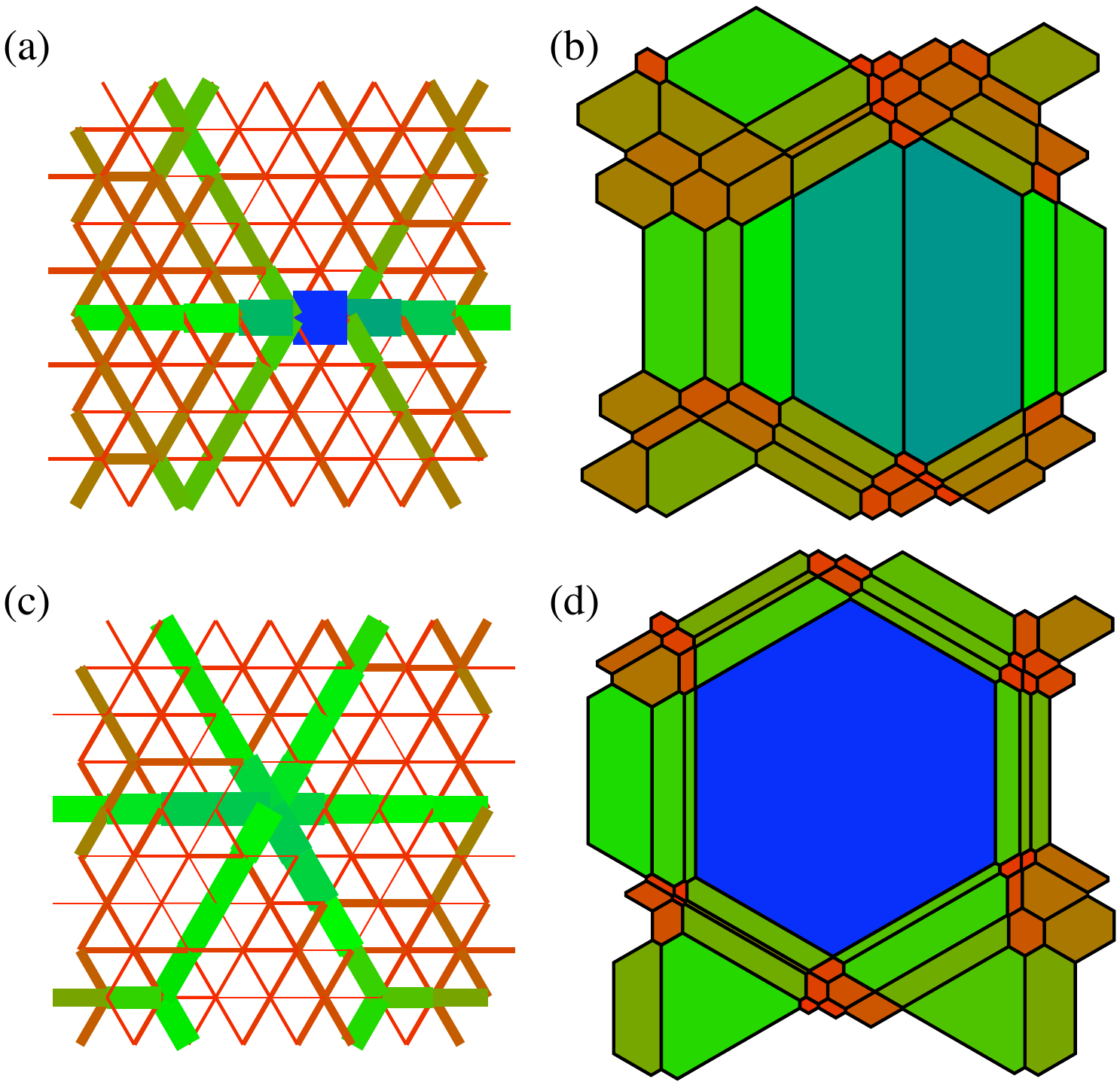}
\caption{(a) Typical force network containing a large contact force. Edge thicknesses are proportional to force magnitudes. (b) Reciprocal tiling of the force network in (a). (c) Typical force network containing a large local pressure (d) and its tiling. Colors map redundantly to force magnitudes or tile areas.}
\label{fig:um2}
\end{figure}

In simulations we employ umbrella sampling, which permits extremely precise determination of the probability density of large stresses \cite{vaneerd07}. Monte Carlo simulations sample force networks ${\bf f}$ with a probability proportional to their statistical weight, i.e. in the
force network ensemble each force network that satisfies local force
balance and the boundary conditions is equally likely. As the vast
majority of force networks only contain contact forces of order of
magnitude $\left\langle f\right\rangle$, large contact forces or local
pressures are hardly sampled and it is not possible to obtain $P(f)$
or $P(p)$ accurately for large $f$ or $p$ respectively.

To improve the sampling for large contact forces or local pressure, we
employ the umbrella sampling method \cite{tor771,fre021}. The central
idea is to create a bias in the force networks obtained in the Monte
Carlo simulations, and to correct exactly for this bias afterwards. To
illustrate this, let us denote the a priori probability of a force
network ${\bf f}$ by $G({\bf f})$. In the force network ensemble
$G({\bf f})$ equals either $0$ or $1$. Monte Carlo simulations of
the force network ensemble generate configurations with a probability
proportional to $G({\bf f})$ so therefore the average of a property
$A$ can be computed from
\begin{equation}
\left\langle A \right\rangle = \frac{\sum_{i=1}^K A({\bf f}_i)}{K} \,,
\end{equation}
in which ${\bf f}_1, {\bf f}_2 , \cdots , {\bf f}_K$ are the force
networks generated by the Monte Carlo scheme. To generate more force
networks with large forces, consider the ensemble $\pi$ in which the a
priori probability of a force network ${\bf f}$ equals
\begin{equation}
G_\pi({\bf f}) = G({\bf f}) \exp [W({\bf f})] \,,
\end{equation}
in which $W({\bf f})$ is an arbitrary function that only depends on
the force network ${\bf f}$. Monte Carlo trial moves from state ${\bf
f}_0$ to state ${\bf f}_n$ in this ensemble are thus accepted with a
probability \cite{met531}
\begin{equation}
{\rm acc}({\bf f}_0 \rightarrow {\bf f}_n) = {\rm
min}\left(1,\frac{G({\bf f}_n)}{G({\bf f}_0)}\exp[W({\bf f}_n) -
W({\bf f}_0)]\right) \,.
\label{eqn:A3}
\end{equation} 
Ensemble averages calculated in the ensemble $\pi$ can easily be 
converted to ensemble averages in the original force network ensemble, 
as
\begin{eqnarray}
\left\langle A \right\rangle &=& \frac{\int d{\bf f}\, G({\bf f})
A({\bf f})}{\int d{\bf f} \,G({\bf f})} \nonumber\\ &=& \frac{\int
d{\bf f} \,G({\bf f}) \exp[W({\bf f})] A({\bf f}) \exp[-W({\bf
f})]}{\int d{\bf f} \,G({\bf f}) \exp[W({\bf f})] \exp[-W({\bf f})]}
\nonumber\\ &=& \frac{\left\langle A({\bf f})\exp[-W({\bf
f})\right\rangle_\pi}{\left\langle \exp[-W({\bf f})]\right\rangle_\pi} \,,
\label{eq:mappingtofne}
\end{eqnarray}
in which we used the shorthand $\left\langle \cdots \right\rangle_\pi$
for averages computed in the ensemble $\pi$. A smart choice of $W({\bf
f})$ will sample many networks with large contact forces, so $P(f)$
can be computed accurately for large $f$. A convenient choice is to
introduce the order parameter $f_{\rm max}({\bf f})$ as the largest
contact force of a force network ${\bf f}$ and to express $W$ as a
function of $f_{\rm max}$ only. The function $W(f_{\rm max}({\bf f}))$
can be chosen such that the probability distribution $P_\pi(f_{\max})$
(computed in the modified ensemble) is approximately flat. As from
Eq. \ref{eq:mappingtofne} follows that
\begin{equation}
P(f_{\rm max}) = {\rm constant} \times P_\pi(f_{\rm
max})\exp[-W(f_{\rm max})] \,,
\end{equation}
it is convenient to iteratively determine $W(f_{\rm max})$ such that
\begin{equation}
W(f_{\rm max}) = -\ln P(f_{\rm max}) \,.
\end{equation}
To illustrate the use of umbrella simulations, in Fig.~\ref{fig:um1}
we have plotted the probability distributions of $P(f)$ and
$P(f_{\max})$ computed with and without umbrella sampling for a
frictionless triangular lattice of $N=1840$ particles. From this
figure it becomes clear that umbrella sampling increases the accuracy
of $P(f)$ for large contact forces by many orders of magnitude.

To compute the propability distribution of local pressures $P(p)$
accurately for large $p$, we employ the same scheme. However, it turns
out configurations containing a single large local pressure are
qualitatively different from configurations containing a single large
contact force. In the former, the large pressure is spread out over
all contacts of the particle, while in the latter only two grains
experience a large contact force. This is shown in
Fig. \ref{fig:um2}. As a consequence, to compute $P(p)$ accurately for
large $p$ a different order parameter is needed. In our simulations to
compute $P(p)$, we choose $W=W(p_{\max}({\bf f}))$ in which
$p_{\max}({\bf f})$ is the maximum local pressure of a force network
${\bf f}$.

\bibliographystyle{apsrev}
\bibliography{tighe}

\end{document}